\PassOptionsToPackage{unicode}{hyperref}
\PassOptionsToPackage{hyphens}{url}
\documentclass[
  10pt,
]{article}
\usepackage{amsmath,amssymb}
\usepackage{setspace}
\usepackage[T1]{fontenc}
\usepackage[utf8]{inputenc}
\usepackage{textcomp}
\usepackage{tgpagella}
\DeclareUnicodeCharacter{2212}{\ensuremath{-}}
\DeclareUnicodeCharacter{00B0}{\ensuremath{^\circ}}
\IfFileExists{upquote.sty}{\usepackage{upquote}}{}
\IfFileExists{microtype.sty}{
  \usepackage[]{microtype}
  \UseMicrotypeSet[protrusion]{basicmath} 
}{}
\makeatletter
\@ifundefined{KOMAClassName}{
  \IfFileExists{parskip.sty}{%
    \usepackage{parskip}
  }{
    \setlength{\parindent}{0pt}
    \setlength{\parskip}{6pt plus 2pt minus 1pt}}
}{
  \KOMAoptions{parskip=half}}
\makeatother
\usepackage{xcolor}
\usepackage[margin=2.2cm]{geometry}
\usepackage{longtable,booktabs,array}
\usepackage{calc} 
\usepackage{etoolbox}
\makeatletter
\patchcmd\longtable{\par}{\if@noskipsec\mbox{}\fi\par}{}{}
\makeatother
\IfFileExists{footnotehyper.sty}{\usepackage{footnotehyper}}{\usepackage{footnote}}
\makesavenoteenv{longtable}
\usepackage{graphicx}
\makeatletter
\def\maxwidth{\ifdim\Gin@nat@width>\linewidth\linewidth\else\Gin@nat@width\fi}
\def\maxheight{\ifdim\Gin@nat@height>\textheight\textheight\else\Gin@nat@height\fi}
\makeatother
\setkeys{Gin}{width=\maxwidth,height=\maxheight,keepaspectratio}
\makeatletter
\def\fps@figure{htbp}
\makeatother
\providecommand{\tightlist}{%
  \setlength{\itemsep}{0pt}\setlength{\parskip}{0pt}}
\usepackage{amsmath}
\usepackage{float}
\usepackage{longtable}
\IfFileExists{bookmark.sty}{\usepackage{bookmark}}{\usepackage{hyperref}}
\IfFileExists{xurl.sty}{\usepackage{xurl}}{} 
\hypersetup{
  pdftitle={Where reconfiguration makes a structure stronger},
  pdfauthor={Iman Mohagheghian},
  hidelinks,
  pdfcreator={LaTeX via pandoc}}

\title{Where reconfiguration makes a structure stronger}
\author{Iman Mohagheghian\\[4pt]\small\normalfont School of Mechanical Engineering Sciences, University of Surrey, Guildford GU2 7XH, United Kingdom\\[2pt]\small\normalfont Correspondence: i.mohagheghian@surrey.ac.uk}
\date{}

\begin{document}
\maketitle

\setstretch{1.15}
\hypertarget{abstract}{%
\subsection{Abstract}\label{abstract}}

Mechanical metamaterials can reconfigure on command and learn,
so adaptation is a central promise of the field. Can reconfiguration make a metamaterial stronger? The answer depends on the comparison. Against the best fixed design of the same mass, the advantage of reconfiguring is often exactly zero. Reconfiguring helps
only if the self-stress the actuators add lowers every member at its limit simultaneously. It vanishes where a member at its limit is one no
self-stress can change, which the geometry fixes, or where those members
are coupled so that lowering one raises another, which depends on the geometry and the loads together. Where neither block applies the gain can be large, as in a twelve-spoke hub with its worst member utilisation reduced by 29.1\%. Whether reconfiguration is worthwhile is
decided by how the self-stress of a structure meets the
members setting its capacity under load, and can be tested in advance.

\hypertarget{introduction}{%
\subsection{Introduction}\label{introduction}}

Mechanical metamaterials can now switch their stiffness and carry
several functions in one body, and they can reconfigure on command or
learn a response through training {[}1--5{]}. The same architected
solids have been shown to govern toughness and damage tolerance
{[}6,7{]}. Adaptation, or reconfiguration, two terms the literature uses for one another, has become a central promise of the field, and
trusses whose members change length are now built and controlled as
robots with hundreds of actuators {[}8{]}. Those demonstrations show what reconfiguration can do and the field has not asked whether reconfiguration can be used to improve strength.

That question is asked here for pin-jointed frameworks, or trusses, in which each member carries axial force only and can change its length as an actuator. Reconfiguration here means the change of load path which the actuators of a structure make by adding a self-stress, with the geometry and the member sizes left as built. Whether such a change makes the structure stronger depends on what it is compared with. A
fixed design built from the same material can use that material for strength by thickening the members which decide its capacity and thinning the rest,
and a design sized in this way is a more demanding reference than a uniform one. The comparison used here holds the material budget fixed and sizes
the best fixed design it can find against the full set of loads. It then asks how far the actuators lower the worst utilisation with the member sizes frozen. The utilisation of a member is the force it carries divided by its capacity (the yield force in tension and the smaller of the yield and the Euler buckling force in compression). The worst utilisation is then the largest value over all members and all load cases. The fraction by which they lower it is called the \emph{premium} of adaptation here, since it is what the actuators add above the best fixed design found for the same material. An adaptive structure can always leave
its actuators idle and so is never worse than the fixed design it starts
from {[}9{]}, which means the sign of the answer carries no information.
The whole question is whether the premium is above zero and by how much.

Neither the metamaterials literature nor the literature on adaptive load-bearing structures has answered this question, although each holds one part of it. The metamaterials literature demonstrates the capability without measuring it, since each result shows a structure adapting and
none asks what the adaptation is worth against a fixed design given the
same material. The adjacent literature on adaptive load-bearing
structures does measure adaptation in deflection and through it in
whole-life energy and mass, and it holds the strength limit as a
constraint which the structure must meet with its actuators switched off
so a power failure costs nothing beyond serviceability {[}10--12{]}.
Whole-life energy or mass is the objective in those studies and the
passive comparator is re-optimised rather than budget-matched, so a
comparison at fixed material does not exist there. The nearest formulation is that of Senatore and Wang {[}13{]}, which optimises topology, member sizes and actuator placement at once and reports the mass saved against a topology-optimised passive structure. There the total material is the quantity minimised and the saving is the result, whereas here the material is held fixed, the passive design is sized first and the quantity sought is the strength the actuators add to it. On the experimental side, Adam and Smith {[}14{]} lowered the
worst-case utilisation of a tensegrity by injecting self-stress at fixed
member sections and measured the result. Their experiment is the closest mechanical precursor, since self-stress was used to lower stresses at fixed member sections, although they had no budget and no fixed competitor. Two further results from the same community are used directly. In a statically determinate structure the actuators change no member force {[}15{]}, so its premium is exactly zero. Steffen and co-workers {[}16,17{]} show that a redundant truss can also contain members whose force no actuator can change. Their analysis is made with no load on the structure, so it does not say whether such a member is one which limits the strength.

Figure 1 asks the question on four architectures ordered by their number of states of self-stress and shows that the premium does not follow that order.
The four-bar pyramid with one state has a premium of 16.7\% and the twelve-spoke hub with nine has a premium of 29.1\%, while the octet with ten states returns exactly
zero and the triangulated lattice with 33 has a premium of 2.6\%.

\includegraphics[width=1\textwidth,height=\textheight]{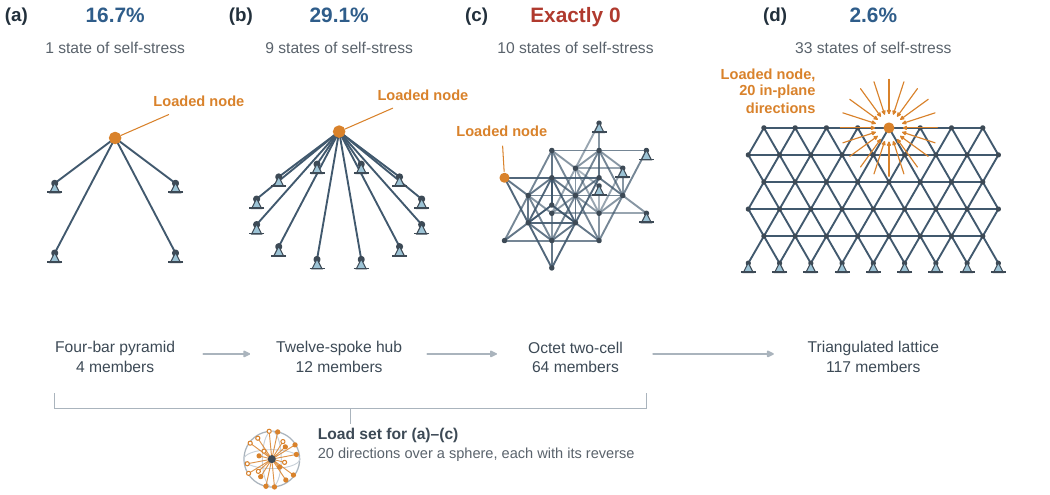}

\textbf{Figure 1.} Four architectures ordered by their number of states
of self-stress, with the premium each one gains above it. The premium is
the reduction in worst utilisation that adaptation achieves
over the best fixed design at the same material, with members limited by Euler buckling and every architecture
built to the same median member slenderness. Orange marks a loaded node
and each node is loaded in one direction at a time. The three 3D
structures in (a)--(c) share one set of twenty directions over a sphere
(bottom), which holds the reverse of every direction it contains and is
listed in Supplementary Table S1, and open circles mark directions
pointing away from the viewer. The lattice in (d) is loaded in twenty
directions spaced evenly round its plane. Redundancy rises from (a) to
(d) but the premium does not follow it, and the octet returns exactly
zero.

A zero premium is already known for statically determinate structures, where actuators produce no internal force {[}15{]}, and what is new here is the metric and the structures in which the zero is found. Three things are reported here. The first is the quantity itself, the premium of adaptation measured against the best fixed design at the same material, which neither literature has measured. The second is a test which says in advance where that premium is exactly zero (stated and proved as a proposition in Methods), which means that the actuators add nothing to the load the best fixed design already carries. The test asks whether any self-stress can lower all the members at the limit together. The third is what the test finds on eight structures loaded in turn at every node. Zeros are common and every one is proved by certificate. Four of these structures span one to 33 states of self-stress and the other four were chosen as the cases most likely to contradict the test , so all eight serve as worked examples of a test which applies to any framework. The test says nothing about the size of a positive premium, which the Results trace to the slenderness of the members and the geometry at the loaded node. Whether reconfiguration makes a structure stronger is therefore settled by the geometry and the loads, and can be determined before the structure is built.

\hypertarget{results}{%
\subsection{Results}\label{results}}

\hypertarget{the-matched-budget-comparison}{%
\subsubsection{The matched-budget
comparison}\label{the-matched-budget-comparison}}

Every result in this paper compares two designs built from the same
material budget. The fixed design is free to size each of its members
for the whole range of loads it will meet, and it spends the budget
where the worst case falls. The adaptive design keeps those member sizes and adds a self-stress through its actuators, which it may choose afresh for each load case, and this is called the retrofit comparison here. The premium is the fraction by which adaptation
lowers the worst utilisation of the fixed design, taken over every load case in the set. Since the adaptive design can always apply zero self-stress it contains the fixed design and the premium is never negative {[}9{]}, so the question is how large it is and where it vanishes. Every structure here carries a unit force at one node in each of twenty directions in turn, spread over the sphere or round the circle (Methods), so the fixed design is sized against all twenty cases while the actuators answer each case on its own.

The size of the premium depends first on what adaptation is compared against, and
Figure 2 shows how much. On the octet two-cell the actuators lower the
worst utilisation of a uniform design by 2.7\%, a small gain which
disappears once the uniform design is replaced by the best fixed design
at the same material. Sizing alone lowers the worst utilisation by
58.9\% and the actuators then add exactly nothing, so the whole of the
apparent gain was available to a passive design which used its material well. The triangulated lattice sits between the two, where sizing
spreads the material by a factor of 76 from the thickest member to the
thinnest and the actuators still find 2.6\%. The twelve-spoke hub is the
opposite case, since its best fixed design stays within 6\% of uniform
and the actuators deliver 29.1\% on top of it, so the hub carries 1.41
times the load its best fixed design carries, a ratio called the load factor here and defined in Methods.

The fair test is therefore the one a designer faces at the design stage,
which is whether to fit actuators or to put the same material into the best fixed design. Every premium that follows is measured against that
design. It is the best that two independent optimisers and twenty further starts could find for each structure, which is evidence and not a proof
that no better design exists (Methods).

\includegraphics[width=1\textwidth,height=\textheight]{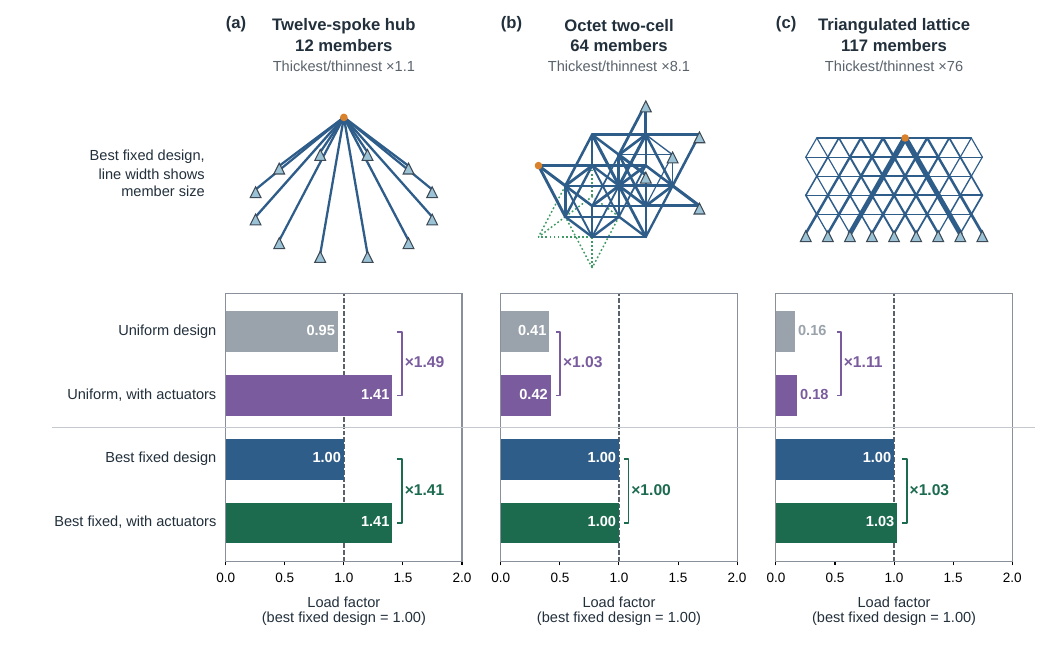}

\textbf{Figure 2.} The fixed competitor decides what adaptation is
worth. Each structure is built from the same material budget in four
ways, and each bar is the load factor of that design, the largest
multiple of the reference load it carries before its worst member
arrives at capacity, with the reference load set by the best fixed
design of that structure so that it reads 1.00 (Methods). Purple
brackets give what actuators gain on the uniform design and green
brackets what they gain on the best fixed design at the same material,
which is the comparison used throughout. Each bracket is the ratio of
the two load factors it joins, which is 1/(1 − premium) where the
premium is the fraction by which the worst utilisation falls, so the
premiums are 32.7\%, 2.7\% and 9.5\% on the uniform designs and 29.1\%,
0 and 2.6\% on the best fixed designs. (a) On the twelve-spoke hub the
best fixed design is close to uniform at 0.95 against 1.00, and the
actuators deliver 29.1\% and take it to 1.41. (b) On the octet the
uniform design carries 0.41 and sizing the members takes it to 1.00,
while the actuators then add exactly nothing, so the 2.7\% they appear
to add on the uniform design is already achieved by sizing alone. (c) On the
triangulated lattice the uniform design carries 0.16, sizing takes it to
1.00, and the actuators still find 2.6\% which brings it to 1.03. Each
structure is drawn as its best fixed design, with line width showing
member size on a scale shared by the three panels, and the ratio of
thickest to thinnest member is given above it. On the octet the sizing
takes nine members which carry no load from this node down to the lower
limit on member area (Methods), drawn as green dots, and the ratio is taken over
the other 55 of its 64 members. Euler member capacity, median slenderness 91, orange marks the loaded node.

\hypertarget{what-the-actuators-do-inside-the-structure}{%
\subsubsection{What the actuators do inside the
structure}\label{what-the-actuators-do-inside-the-structure}}

Figure 2 gives the premium as a single number for each structure, and
Figure 3 opens it member by member. Each row follows one load case, the one which remains the worst after the best actuation . The member sizes are those of the best fixed design and do not change, so the only thing which differs between the panels of a row is the force the actuators add. That added force is a self-stress, a set of member forces which balances itself without any external load, and the actuators
may choose a different combination of the available self-stresses for
every load case.

In every case the self-stress rearranges the member forces so that more members arrive at the limit together, which draws on capacity the fixed design left unused. The worst utilisation falls only where that rearrangement brings the governing members down with the rest. On the hub the actuators put tension into the two spokes which govern the fixed design, that is the two at its worst utilisation, and move compression into spokes which had capacity to spare, so the two governing spokes come down while eight others rise to meet them and the number of members at the limit grows from two to ten. The worst utilisation falls by 29.1\% and the load factor of the hub rises to 1.41. The lattice shows the same pattern at a smaller scale, where the number of members at the limit grows from 14 to 41 and the worst value falls by 2.6\%.

The octet shows what happens when this levelling is not possible. The actuators can still bring ten members to the limit instead of three, yet the worst value does not move, because no combination of self-stresses lowers it.

The premium therefore depends on which members are at the limit together
and on whether the self-stress the actuators add can relieve all of them
at the same time. The next section turns that condition into a test
which says in advance where the premium is exactly zero.

\includegraphics[width=1\textwidth,height=\textheight]{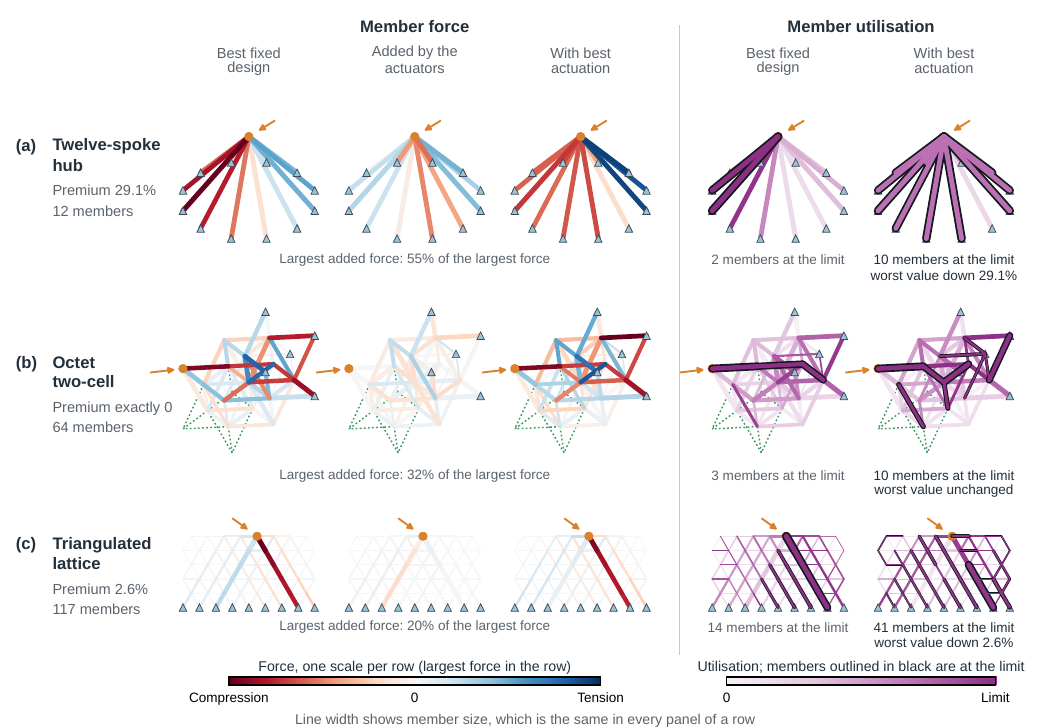}

\textbf{Figure 3.} What the actuators do inside the structure. One row per structure, and each row shows a single load case, the one that remains the worst after the best actuation. Member
sizes are frozen, so line width, which shows member size, is the same in
every panel of a row, and the nine octet members at the lower limit on
member area are drawn as green dots.
Left, member force from compression (red) to tension (blue) on one scale
per row: the best fixed design, the force added by the actuators, which
is a self-stress, and their sum, the design with the best actuation.
Right, member utilisation from zero to the worst value of the best
fixed design, with every member at the limit outlined in black. (a) On
the twelve-spoke hub the actuators add tension to the governing spokes,
10 members end at a common limit, the worst value falls by 29.1\% and
the load factor rises to 1.41. (b) On the octet no self-stress lowers
the worst value, so the added forces drawn, up to 32\% of the largest
force, are one of many equally good choices and applying none would do
as well; the number of members at the limit rises from 3 to 10 but the worst value does not fall and the load factor stays at 1.00. (c) On the triangulated
lattice the added forces are small, the number of members at the limit rises from 14 to 41, the worst value falls by 2.6\% and the load factor rises to 1.03.
Best actuation is defined in Methods, and where several combinations
give the same lowest value the one returned by the optimiser is drawn.
Orange marks the loaded node and the arrow the load direction. Euler
member capacity, median slenderness 91.

\hypertarget{exact-zeros-the-self-stress-test}{%
\subsubsection{Exact zeros: the self-stress
test}\label{exact-zeros-the-self-stress-test}}

The actuators can only add a self-stress, and that self-stress may
combine the independent self-stress states of the structure in any
proportion. The worst utilisation of a load case therefore falls only if
one such combination lowers every member which stands at the limit in
that case, and lowers them all together. If a single governing load case, meaning a case in which the worst utilisation of the design is attained , has no such combination, the premium is exactly zero. More actuators, or
actuators placed elsewhere, cannot change this, because they can only
produce combinations of the same self-stress states. Figure 4a sets out
this test, and for each governing load case a linear programme searches
every combination of self-stress states for one that lowers all the
members at the limit (Methods). The test returns a zero in two ways and
the octet shows both.

The first block is a member which no self-stress can change. How strongly the self-stress acts on a member is measured by its \emph{reach}, which lies between 0 and 1 and is defined in Methods (Eq. (4)). A self-stress of unit size can change the
force in a member by at most the square root of its reach, so a member
of zero reach carries the same force whatever the actuators do. Supplementary Figure S1 maps reach for the three structures of Figure 3,
where every spoke of the hub has a reach of 0.75 while the octet runs
from 0.42 down to zero. Reach is known in structural mechanics as the
static indeterminacy factor of a member {[}18{]}. What is new here is where it is read, which is at the members standing at the limit in the load case that governs, since a member of zero reach matters only when it is one of them. Twelve of the 64
members of the octet two-cell have zero reach, all of them at its free
end away from the supports (dashed in Figure 4b). Figure 4b shows the octet loaded at the corner node of its free end, under the governing load case drawn in Figure 3b. In that case the best fixed design brings three members to the limit together, and one of the three is among the twelve members of zero reach. The actuators can lower the other two members without limit and the worst utilisation still stays at the value the unreachable member sets, so the load factor stays at 1.00 and the premium is zero. Reach is fixed by the geometry of the structure, which includes where the nodes sit, how the members join them and which nodes are supported, so it can be mapped before any load is applied and does not change with slenderness or material. Whether a member of zero reach is at the
limit does depend on the loads, so the premium of the same octet can vanish at one
loaded node and not at another. The same block occurs in a flat lattice of a different kind (Supplementary Figure S2), so it is not special to the octet.

The second block needs no such member. Loaded alone at the node marked in Figure 4c, the octet brings three members to the limit, and all three can be reached,
yet two of them are coupled so that every self-stress which lowers one raises the other (Figure 4c). Supplementary Figure S3 shows the same for every pair of the three, so the coupling sits only in that one pair. The set of everything the
actuators can do then misses the region where both fall, and the premium
is again exactly zero. Every self-stress changes the force in E and F by the same amount, and since one of them is in tension and the other in compression, one cannot fall without the other rising.  Both zeros occur in this example of a statically indeterminate structure, which has ten independent self-stresses and an actuator on every member. This shows that redundancy alone does not make the premium positive, and it is different from a statically determinate structure, where a zero follows directly from the absence of any self-stress for the actuators to act on.

\includegraphics[width=1\textwidth,height=\textheight]{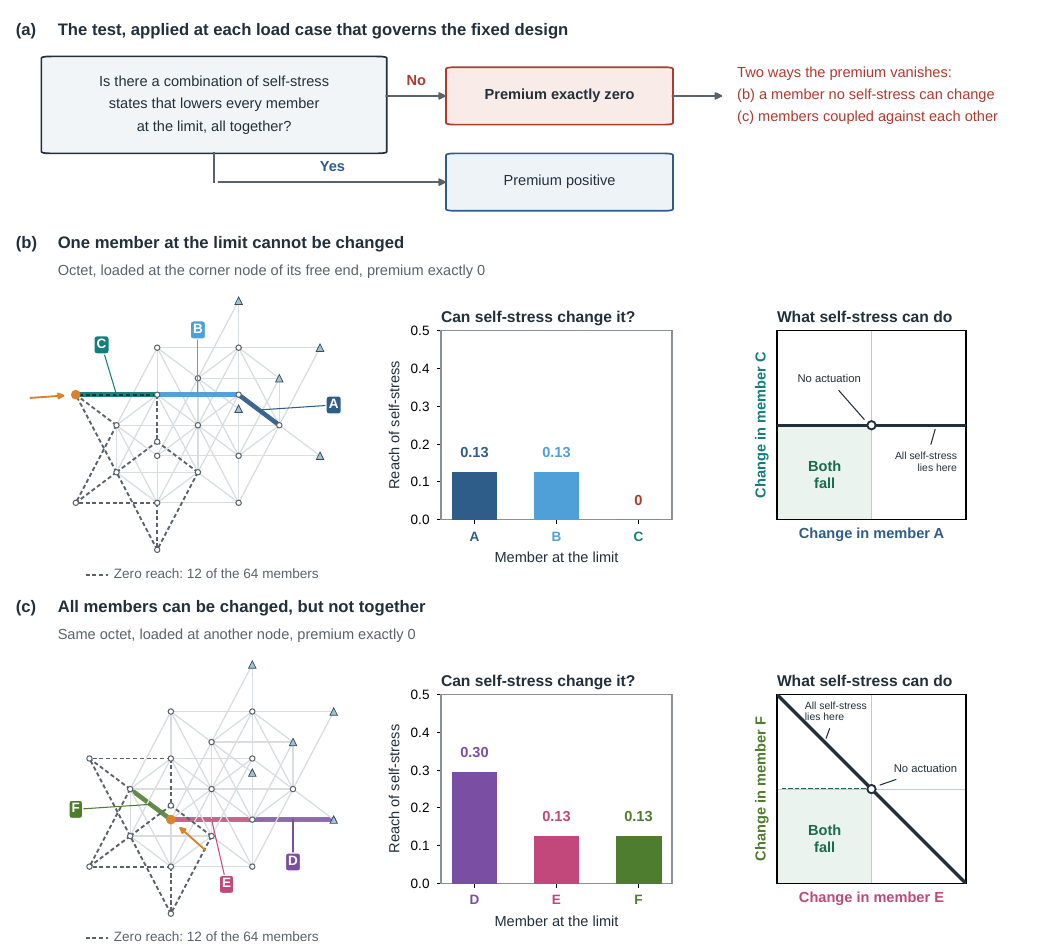}

\textbf{Figure 4.} Why the premium is exactly zero, and the two ways it
happens on one octet. (a) The test. At each load case which governs the
best fixed design, the premium can be positive only if some combination
of self-stress states lowers every member at the limit all together;
where none does, the premium is exactly zero. (b) The octet loaded at the corner node of its free end, in orange, under one of the load cases which govern its best fixed design. Three members are at the limit, lettered and
coloured as on the axes, and member C has zero reach, meaning no
self-stress in the structure changes its force, so the line of
everything self-stress can do is flat in member C and never enters the
quarter where both members fall. Twelve of the octet's 64 members share
that property and are drawn dashed in both rows, all of them at the free
end of the two-cell, away from the supports. (c) The same octet loaded
alone at the node in orange, a different node from (b). All three members at the limit can be changed and none
is among the dashed members of zero reach, but members E and F are
coupled so that whatever lowers one raises the other, and the line again
misses the quarter where both fall. Reach is defined in Methods (Eq. (4)),
and the largest value in the octet is 0.42. Whether a member of zero
reach is at the limit does depend on the loads. The right-hand panels carry no scale because any multiple of a self-stress is also a self-stress, so only the direction of the line matters. A linear programme over the whole set of members at the
limit confirms both zeros. Euler member capacity, slenderness 91, twenty
load directions at the loaded node.

\hypertarget{geometry-at-the-loaded-node}{%
\subsubsection{Geometry at the loaded node sets the value}\label{geometry-at-the-loaded-node}}

Both blocks of the self-stress test depend on the geometry of the structure and on the loads it carries. A designer cannot usually change the loads but can change the geometry, so Figure 5 asks what a change of geometry at the loaded node does to the premium. The test is made on the triangulated lattice of Figure 2, chosen because its premium of 2.62\% is small without vanishing, so a change of geometry at the loaded node can be seen to move it either way. The plain lattice is the reference, so its load factor as a fixed design reads 1.00 and it carries 1.03 times that with the best actuation (Figure 5a). Four further configurations are built from it by adding members at the loaded node. That is where the load enters the structure and where the members which decide its capacity sit. New members there change both which members are at the limit and what the self-stress can do to them. Every configuration is built from the same amount of material as the plain lattice, so the material for the added members comes from thinning the others, and each configuration is sized and actuated against its own best fixed design.

In the first configuration seven new bars run straight from the loaded node to the supports, each a single member, and it was found to be the worst of the five (Figure 5b). Each added member spans the width of the lattice in one piece and buckles at a fraction of the force a chain of short members carries, so the configuration carries 0.12 of the reference load as a fixed design and 0.14 with actuation,
about eight times weaker than the plain lattice. Its premium of 16.69\% is the highest of the five, yet it means only that actuation lifts the configuration from 0.12 to 0.14 of the reference load, and both values sit far below the plain lattice at 1.00. A large premium can therefore sit on a design which is worse than the plain lattice it was built from, which is why every premium in this paper is reported beside the load
factor it is measured from.

In the second configuration the same seven bars are jointed to the lattice wherever they cross it (Figure 5c). Each crossing becomes a node, so every bar becomes a chain of short members and the lattice members it crosses are split as well, which takes the member count from 124 to 238 at the same material. That reverses the result. The premium rises to 9.77\% and the load factor
rises to 1.14 as a fixed design and 1.27 with actuation, the highest of
the five, so this design carries 23\% more load than the plain lattice
does with its own best actuation. The gain is therefore real in both
senses, since the adaptive design improves more and the structure it
improves is stronger to begin with.

Every added bar starts at the loaded node, and where its other end is connected matters as much as whether it is jointed to the lattice along the way. In the third and fourth configurations the seven bars run from the loaded node only to its nearest neighbours in the lattice rather than to the supports, left crossing the lattice without joints as in (b) (Figure 5d) and jointed at every crossing as in (c) (Figure 5e). They carry 0.67 and 1.08 of the reference load with actuation, so bars which connect the loaded node only to its neighbours leave the structure weaker when unjointed and 5\% stronger when jointed, against 23\% for the jointed bars to the supports. What sets the jointed long bars apart is that they connect the loaded node to the supports by a jointed path, which is the shortest route the force can take, and that path carries the load while spreading the same material over a total member length of 152.3 against the plain lattice's 117.0, so its members are thinner on average. Figure 5f sets the five side by
side.

The five configurations also show what kind of change matters. Adding bars at the loaded node alters which members are at the limit and what the self-stress can do to them, and the premium moves with both. The geometry at the loaded node therefore sets the size of the premium as well as whether it is zero.
Supplementary Figure S4 gives the material, the reach and the member
utilisation before and after actuation for all five configurations.

\includegraphics[width=1\textwidth,height=\textheight]{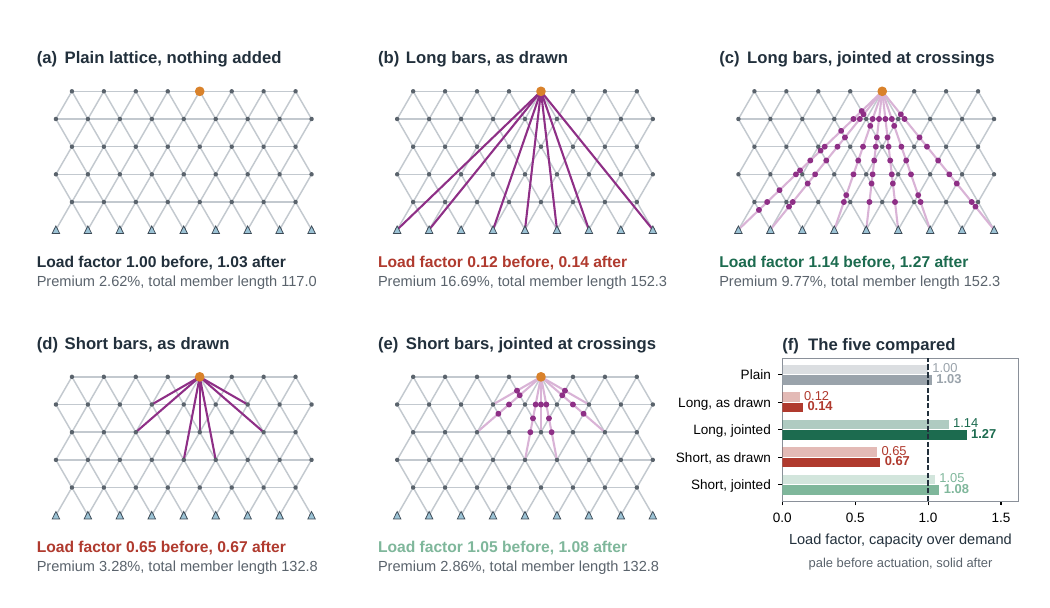}

\textbf{Figure 5.} How the geometry at the loaded node changes the premium. A triangulated lattice of 51 nodes and 117 members is
loaded at the middle of its top row in twenty in-plane directions, and
every configuration is built from the same material budget and measured
against its own best fixed design, with the buckling constant held at
81.9 for all five.
Seven bars are added at the loaded node in (b) to (e), each a single member where it is unjointed and a chain of members where it is jointed, drawn in full plum where they cross the lattice without a joint and in pale plum with
their new nodes marked where they are jointed at every crossing. Under
each panel is the load factor of the design before and after the best
actuation, the premium it gains over its own best fixed design, and the total member length, which is not a measure of material since the budget is the same in every panel. The reference load is the largest load the plain lattice carries
as a fixed design, so (a) reads 1.00 before actuation (Methods). (a) The
plain lattice carries 1.03 with the best actuation, a premium of 2.62\%.
(b) Long bars running to the supports carry 0.12 before actuation and
0.14 after, because each one is a single long member which buckles, and the premium of 16.69\% is the largest of the five. (c) The same bars jointed where they cross carry 1.14 before actuation and
1.27 after, with a premium of 9.77\%. (d) Short bars to the nearest nodes carry 0.67 with actuation and (e) the same bars jointed carry 1.08, so connecting the loaded node to its neighbours does not secure the gain. (f) The five
compared, pale before actuation and solid after, with the dashed bar at
the reference load, red where a design ends below the plain lattice and
green for the best of the five. Euler member capacity and matched material budget, with member sizes free in the fixed design and frozen under actuation.

\hypertarget{where-the-premium-does-not-vanish}{%
\subsubsection{Magnitude where the premium does not
vanish}\label{where-the-premium-does-not-vanish}}

Where neither block is present the test says only that the premium is
positive, and Figure 6 asks how large it is and what it depends on. The truss model used throughout the paper contains a single free parameter, the slenderness of the median member, which sets how many members are limited by buckling rather than yield (Methods). This parameter is varied here. Slenderness is
expressed as the ratio of member length to radius of gyration that
design codes are written in. The three structures of Figure 1 with a positive premium are swept from the stocky end to well beyond the limit those
codes impose, and the geometry and the material budget are held exactly
as they are elsewhere in the paper, as is the load set.

Each of the three premiums is flat at the slender end of the sweep. The
hub lowers its worst utilisation by 32.54\% for every slenderness
above about 170 and carries 1.48 times the load of its best fixed
design, while the pyramid lowers it by 16.68\% and carries 1.20 times, and the
lattice by 4.21\% and carries 1.04 times. On that plateau every
member which decides either design is in compression and limited by
buckling (Supplementary Figure S5), so a change of slenderness scales every capacity that matters by the same factor and the ratio between the two designs is unchanged.

The descent below the plateau is set by which limit binds the worst
member. A member in tension is limited by yield whatever its
slenderness, so as the members get stockier the worst member of a design
passes from a buckling-limited strut to a yield-limited tie. Once it has
done so the premium no longer depends on slenderness at all. The pyramid
has four members and its worst member changes over between slenderness
91 and 62, so its curve drops to 11.20\% there and is flat again. The hub has twelve
members carrying twenty directions and the changeover is spread across
them, and four of the ten members at its limit are still buckling-limited
at the stocky end of the sweep (Supplementary Figure S5), so its premium
slides to 18.30\% and is still falling where the sweep stops. The lattice slides in the same way over its 117 members, from 4.21\% on the plateau to 0.16\% at the stocky end, and is still falling where the sweep stops. The octet is exactly
zero at all fourteen slendernesses tested, which the self-stress test
requires, because a member of zero reach has zero reach whatever its
capacity. A premium is therefore one point on a curve set by the material, while a zero caused by a member of zero reach does not move along that axis, so the self-stress test is a property of geometry and loads and the magnitude it permits is a property of the material as well.

The plateau is where the finite-element check is made. The hub of Figure
6 at a slenderness of 653 was built in Abaqus as a truss with the member
areas and the twenty load directions of the reduced model under the same
buckling capacity (Methods), and the retrofit self-stress was imposed as
an initial stress in each member. Abaqus returns the same premium of
32.54\%, and the worst utilisations of the fixed and of the actuated
design agree with the reduced model to within 0.011\%. The agreement is a check of the reduced model,
which is linear and elastic, and it says nothing about what happens once
a member buckles.

\includegraphics[width=1\textwidth,height=\textheight]{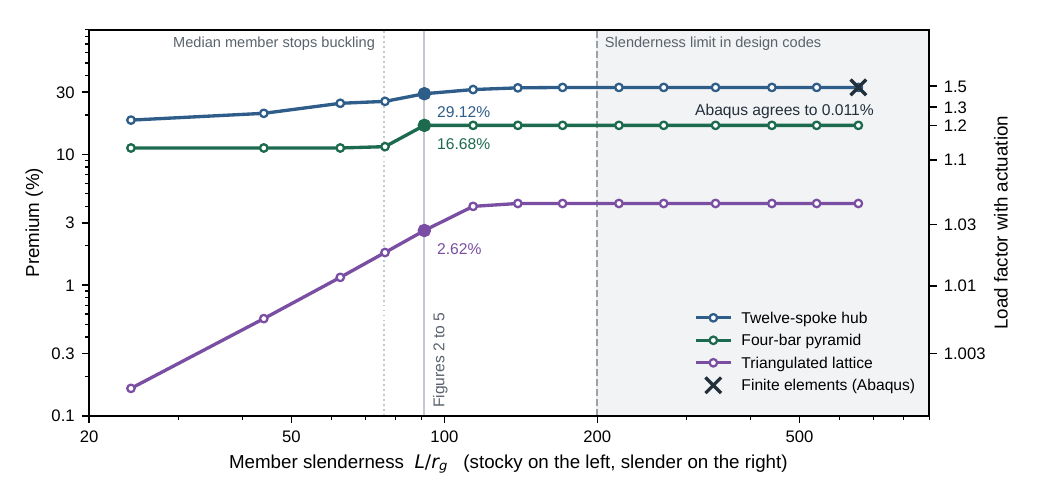}

\textbf{Figure 6.} The premium against member slenderness for the three
structures of Figure 1 with a positive premium. The premium (left axis) and the
load factor with actuation (right axis, Methods) are plotted against the
slenderness ratio of the members, with the matched material budget, the
retrofit actuation and the twenty-direction load set held as everywhere
else in the paper. Filled markers and the grey line mark the slenderness
used in Figures 2 to 5. Above the slenderness limit that design codes
impose (shaded) every premium is flat, and at the stocky end the pyramid's premium is set by yield alone while the hub and the lattice are still falling where the sweep stops. The cross is the finite-element result for the hub,
32.536\% in Abaqus against 32.543\% from the reduced model on the same
twenty load cases. The octet is exactly zero at all fourteen
slendernesses tested and is not drawn.

\hypertarget{the-test-at-every-loaded-node}{%
\subsubsection{The test at every loaded node of eight structures}\label{the-test-at-every-loaded-node}}

The sections so far have shown the self-stress test on single load cases. This section checks it against every loaded node of eight structures. The test was fixed before any further run, then every unsupported node of eight structures was loaded in turn and the test's verdict, zero or positive, was compared with the measured premium at each of the 124 nodes, and every zero was then proved rather than measured. Four of the structures are those of Figure 1 and four were added to test the prediction under conditions the first four do not cover.
The staggered grid is a second flat lattice with fewer members than the triangulated one (48 against 117) and a different member pattern. The barrel vault and the dome are curved lattices. The spoked grid is the staggered grid with six extra members running from the loaded node straight to the supports, the change which Figure 5 made on the triangulated lattice.
The eight structures have 124 unsupported nodes between them, from one on the
pyramid to 42 on the triangulated lattice, and each node is loaded in
turn in its own twenty directions and sized and actuated afresh against
its own best fixed design (Figure 7a, d).

At every one of the 124 nodes the self-stress test predicts whether the
measured premium is zero or positive. Where the test finds no
self-stress that lowers every member at the limit the premium is exactly
zero, and where it finds one the premium is above zero (Figure 7b).
Of the 124 nodes, 31 are exact zeros and 93 are positive. The zeros are not
spread evenly (Figure 7c). Sixteen of the eighteen octet nodes are zero
and so are six of the seventeen dome nodes and six of the nine barrel
nodes, while the staggered grid has two zeros in eighteen and the spoked grid one,
and every node of the triangulated lattice has a positive premium. The test decides whether the premium is zero and says nothing about how large a positive premium is, which across the eight structures runs from 0.04\% to 49\%.

Figure 8 shows why the octet is zero at so many nodes when the actuators
can do so much. At every loaded node the actuator lowers the most loaded member in most of the twenty directions and often by more than a tenth, yet
sixteen of the eighteen nodes have a premium of exactly zero. A design is set by the directions which push it to its limit, which on the octet are most of the twenty, and the premium is measured on the worst of them after actuation. At each of these sixteen nodes at least one of those directions carries a block, so it stays at the limit whatever the actuators do to the others, and the worst case after actuation is no better than before. The premium of the node is therefore exactly zero even where most directions can be lowered. The two nodes with a positive
premium are the two where no governing direction carries a block, and they
gain 0.84\% and 0.09\%. Adaptation helps most load cases on the octet
and gains nothing, because the load cases it cannot help are the ones
which decide the design.

The zero directions above come from the optimiser which searches for the best actuation, and a zero from that search could mean only that it stopped without finding a better design. To rule that out, every one of the 31 zeros of the sweep (Figure 7b) was proved with the certificate of Eq. (7) (Methods), which establishes that no self-stress can lower the members at the limit, so that no better actuation exists. A certificate was found at each of the 31 and none exists at any of the 93 positive nodes.

The 31 zeros fall into the two kinds the test allows. Six of them come from a member of zero reach at the limit, four on
the octet and one each on the staggered grid and the spoked grid.
The other 25, which have no such member at the limit and include every
zero on the dome and the barrel vault, are coupled zeros, so they hold for the
twenty directions of Table S1 and could change under another load set.
At the two nodes of Supplementary Figure S6 the coupling is a pair of members meeting at the loaded
node which respond in exact opposition to every self-stress, and on the barrel vault every certificate is such a pair.

The two kinds of zero differ in how they respond to slenderness. The check was repeated with each of the eighteen unsupported nodes of the octet as the loaded node in turn, at five slendernesses from 24 to 653. A member of zero reach stays unreachable at every slenderness, since reach does not depend on capacity, and every zero caused by one at slenderness 91 stayed at zero at all five. Coupling is less robust, because slenderness changes which members share the limit and how their capacities weigh them. At one node the octet carries a premium of 1.4\% at slenderness 171 and above, and at slenderness 91 and below the premium is exactly zero with no member of zero reach at the limit, because the governing load direction itself changes with slenderness and the case which governs at 91 has a coupled pair at the limit (Supplementary Figure S7). Eleven of the eighteen nodes change between zero and positive in this way. A zero of the first kind therefore survived every change of slenderness tested, while a zero of the second kind has to be checked at the slenderness and loads of the design.

The zeros do not depend on how much force the actuators can add or on how many members carry one. Bounding the force each
actuator may add, or actuating the members one at a time, leaves the
octet and the barrel vault at exactly zero at every bound and every count (Supplementary Figure S8). This is what the test predicts, since a zero caused by the pattern of the self-stress cannot be recovered with more actuator force or more actuators. The hub attains its full premium only when the
force shift allowed is about the member's own working force and only once nine of its twelve members are actuated, and the lattice attains 1.53 \% of its 2.62\% with twelve of its 117 members actuated, so the
size of a positive premium is set by the actuator as much as by the
structure while the zero is set by the structure alone.

Figure 9 reports, for the eight structures of Figure 7 each loaded at the one node marked in Figure 9a, whether a certificate exists. These eight nodes are a subset of the 124 of the sweep, one per structure, chosen so that the certificates can be examined in detail, and every one of the 31 zeros in the sweep carries a certificate of the same kind. At that node the octet and the barrel vault each have a governing load direction which carries a certificate, so those two are zero while the other six have no certified direction and a positive premium (Figure 9b). The certificates settle which nodes are zero, which leaves the question of whether reach, the quantity the test is built on, also says how large the premium is where it is positive. The answer is no (Figure 9c). The barrel vault is
certified zero with a smallest reach of 0.083 at the limit while the
dome has a premium of 2.76\% with a reach of 0.013, the spoked grid 16.9\%
with a reach of 0.108 and the pyramid 16.7\% with a reach of 0.250.

\includegraphics[width=1\textwidth,height=\textheight]{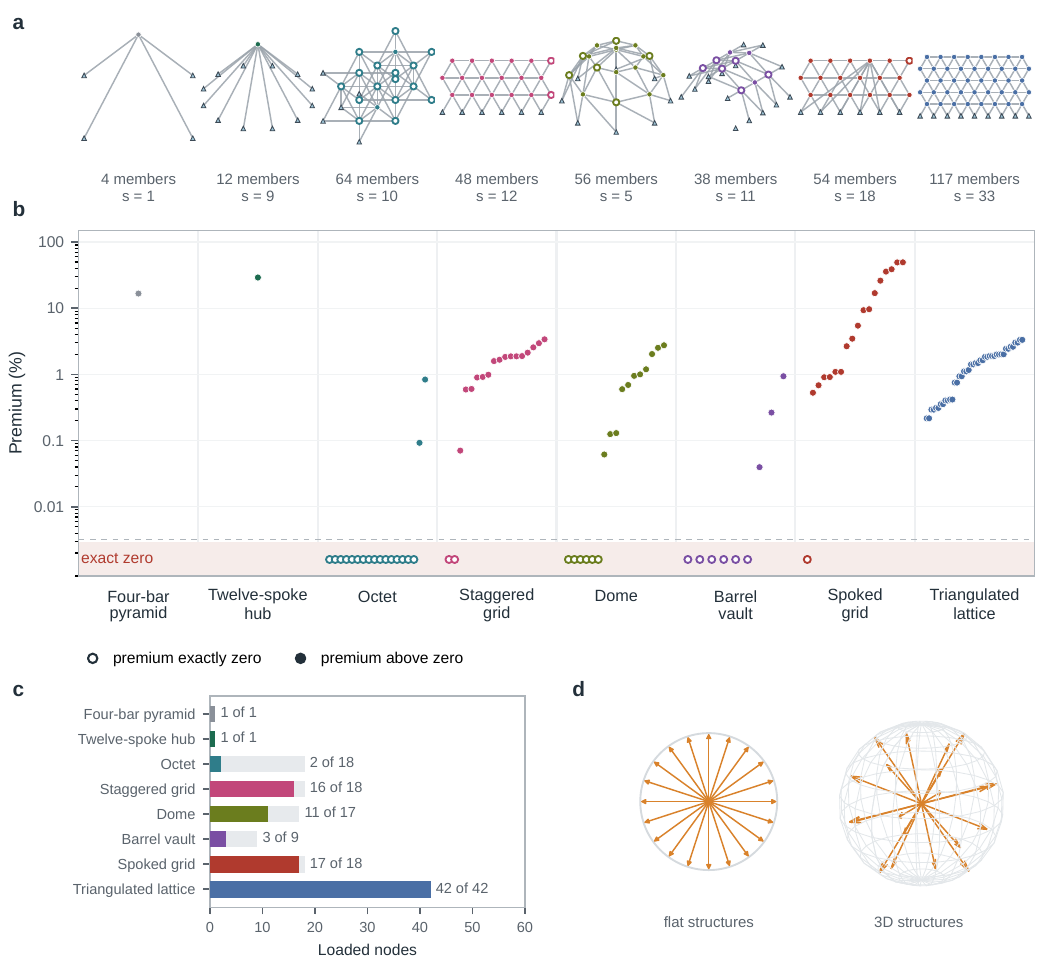}

\textbf{Figure 7.} The self-stress test checked at every loaded node of
eight structures. (a) The eight structures with every loaded node marked
by the same symbol it carries in (b), and beneath each the number of members and of states of self-stress \(s\). (b) One mark per loaded node,
ranked by premium inside each column. An open ring is a node where the test
says the premium vanishes, because no self-stress lowers every member at the limit,
and a filled dot one it does not. Every open ring sits on the exact-zero
band and every filled dot above it, so the test and the measured premium
agree at all 124 nodes. (c) The number of loaded nodes in each structure
with a positive premium, out of the number tested. (d) The twenty load
directions applied in turn at every node, evenly round the circle for
the flat structures and spread over the sphere for the three-dimensional
ones, each direction with its exact reverse. Matched material budget,
retrofit actuation and Euler capacity at the slenderness of Figures 2 to
5 throughout.

\includegraphics[width=1\textwidth,height=\textheight]{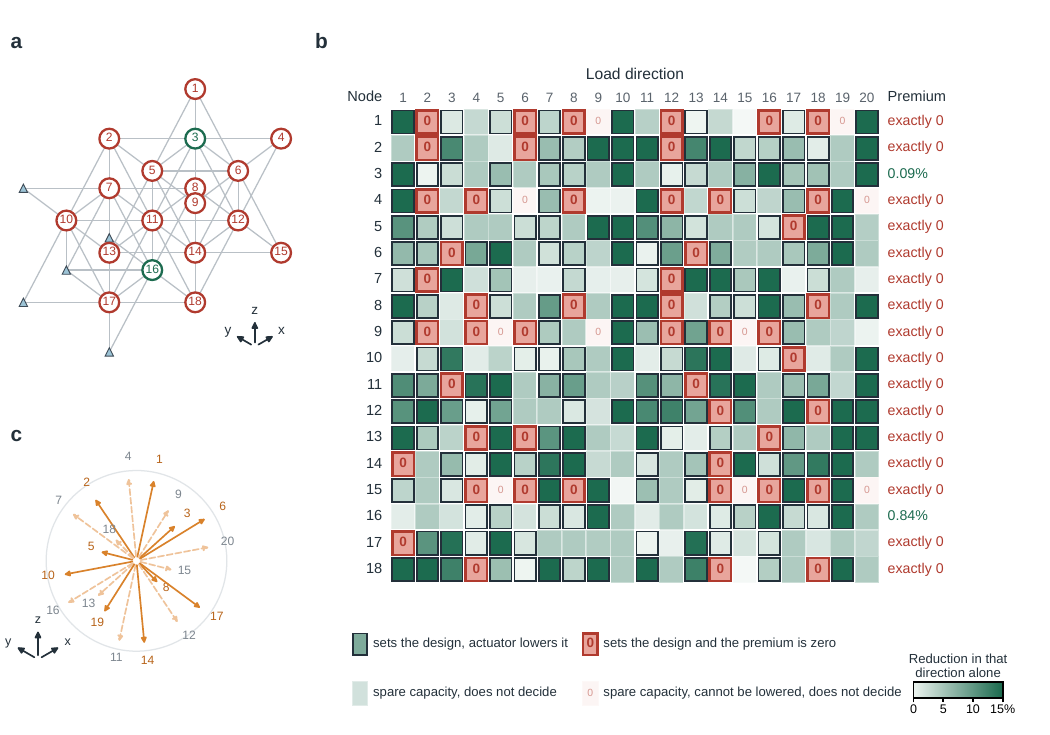}

\textbf{Figure 8.} Which load cases adaptation can help, on the octet at
every loaded node and every load direction. (a) The octet with its
eighteen loaded nodes, numbered in reading order, ringed red where the
premium at that node is exactly zero and green where it is positive. (b) One row per loaded node and one column per load direction, so each cell is one load case and reports the member most loaded under it. Each design
is sized to survive all twenty directions, and the directions which push
it to its limit are drawn at full strength with a frame while the rest
are faded because the design carries them with spare capacity. The shade is how much the actuator lowers that most loaded member on its own. A framed red zero is a direction which sets the design and whose most loaded member cannot be lowered, and one such direction makes the premium of the node exactly zero. (c) The
twenty directions in the same view as (a), solid where they point
towards the viewer and dashed where they point away; each direction and
its exact reverse are ten apart in the numbering. Every node is sized against its own best fixed design at the matched material budget, with Euler member capacity at slenderness 91 and the member sizes frozen under actuation.

\includegraphics[width=1\textwidth,height=\textheight]{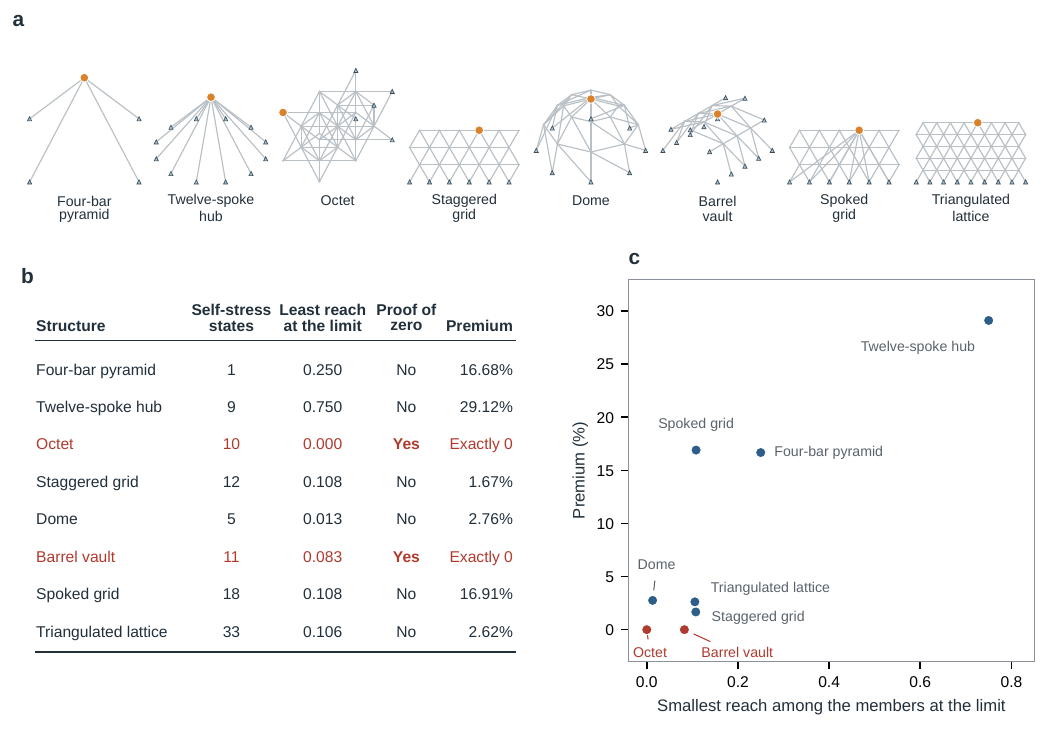}

\textbf{Figure 9.} (a) The eight structures, each with its one loaded node in orange, loaded in turn in twenty directions. (b) For that node, the number of states of
self-stress, the smallest reach among the members at the limit, whether a
certificate of zero exists (Methods), and the measured premium. Rows in
red carry a certificate and their premium is exactly zero, and every
other row has no certificate and a positive premium. (c) The premium
against the smallest reach at the limit for the same eight nodes, red
where a certificate exists. Reach above zero is necessary for a premium and its size does not order the premium. Every node is sized against its own best fixed design at the matched material budget, with Euler member capacity at slenderness 91 and the member sizes frozen under actuation.

\hypertarget{discussion}{%
\subsection{Discussion}\label{discussion}}

This paper gives a test which says in advance where adaptation adds nothing against a fixed design built from the same material. The test says only whether the premium is zero, and where it is not, the paper shows that the geometry at the loaded node sets its size. The test examines the members at the limit of a load case and asks whether any self-stress lowers them all together. It returns a zero in two ways, at a member which no self-stress can change, which depends on the geometry alone, or at members coupled so that no self-stress lowers them together, which depends on the geometry and the loads together. At every one of 124 loaded nodes on eight structures the prediction matched whether the premium was zero or positive, and all 31 zeros are proved by certificate.

The zero itself is not new and the earlier work it comes from should be named. The Kagome lattice was chosen as the actuation lattice two
decades ago because a single member of a pin-jointed repetitive truss
can be actuated without resistance {[}19{]}, and the active-lattice
programme engineered its planar and double-layer Kagome grids to be
statically and kinematically determinate so that no self-stress would
resist the actuators {[}20{]}. In a determinate lattice an actuator generates no internal force, so it cannot lower a worst utilisation. The premium of a determinate structure is therefore zero by definition, and this has already been stated for minimum-weight design {[}15{]}. A member whose force no actuation can change is also
already known, as a zero diagonal entry of the redundancy matrix
{[}21--23{]}, and Senatore and Reksowardojo call an actuation which
changes no member force `impotent' {[}24{]}.

What is new is the metric and the structures the zero is found in. The premium here is a reduction in worst-case utilisation over a load set at matched material, against a fixed design re-optimised on the same loads. The zero appears in five statically indeterminate structures with between five and eighteen states of self-stress and an actuator on every member, and the count of self-stress states does not predict whether the premium is zero. What does predict it is the test, which says for which loaded nodes of an indeterminate structure actuation can make it stronger at all.

The result sits alongside the work on adaptive load-bearing structures
and does not contradict it. The comparison here is a retrofit one, in which the fixed design is sized first and the actuators are added afterwards. The savings reported in that literature are measured on deflection and energy, where
a fixed allocation of material can scale a displacement field and cannot
cancel one. There is then nothing for a fixed design to pre-empt and no
reason to expect the zeros reported here.

The comparison also says something about service life. The actuators can
always be left idle, so in every load case the adaptive structure is at
least as far from its limit as the fixed design, and not only in the
worst case. Any load which breaks the adaptive structure therefore
breaks the fixed design too, so over any sequence of loads the adaptive
structure fails no earlier, whatever the order of the loads and however they are related (Methods). This assumes that the actuators apply the best self-stress for each load before any member is loaded to its capacity. This treats the capacity as fixed when the
structure is built and ends its service life at the first load which
exceeds it, which is the first-passage view of structural reliability
{[}25{]}.

The result has four limits, and the first is that the magnitude is open,
since the test says whether the premium is zero and nothing about its
size, which runs from 0.04\% on the barrel vault to 49\% on the spoked grid. Neither reach (Figure 9c) nor the number of
states of self-stress (Figure 1) orders it. The compensability index and
efficacy measure of Senatore, Virgili and Blandini {[}26{]} are built
for placing actuators on a given design with no matched fixed
competitor, so whether either predicts the premium is left open. Second,
the fixed design is sized by a local method, so each positive premium
is measured against that design and a better one would change it, while
each zero is proved at the design in use. Third, the lifetime account holds only where a
survived load leaves the structure as built, which is the strength
regime, and where fatigue governs, the loading history is not erased.
Two studies show the difference. Efinger and co-workers found that thinning sections to save material shortened the fatigue life of the critical notches, so that an adaptive high-rise gained no service life over its passive counterpart {[}27{]}. Their later bridge study adds a tensioning system to an existing bridge, which slows further fatigue damage but cannot remove the damage already accrued {[}28{]}. Fourth, every number
here is computational and quasi-static on elastic members with a
threshold capacity, and one hub was confirmed in finite elements with no
prototype built. What this paper claims as new is the quantity it
measures, a worst-case strength premium at matched material, and the
test which says where that premium is zero.

\hypertarget{methods}{%
\subsection{Methods}\label{methods}}

\textbf{Self-stress states.} For a pin-jointed structure with \(m\) members, let \(\mathbf{B}\) be the equilibrium matrix which maps the member forces \(\mathbf{t}\) to nodal forces at the free degrees of freedom, so that equilibrium under an applied load \(\mathbf{p}\) reads

\[
\mathbf{B}\,\mathbf{t} = \mathbf{p} \tag{1}
\]

A self-stress is a set of member forces in equilibrium with no load, and
these form the null space of \(\mathbf{B}\), whose dimension \(s\) is the number of
independent self-stress states {[}29,30{]}. An orthonormal basis of this
null space is taken from the singular value decomposition of \(\mathbf{B}\) {[}31{]} and collected as the columns of an \(m \times s\) matrix \(\mathbf{N}\), so that

\[
\mathbf{B}\,\mathbf{N} = \mathbf{0}, \qquad \mathbf{N}^{\mathsf{T}} \mathbf{N} = \mathbf{I}_s \tag{2}
\]

where \(\mathbf{I}_s\) is the identity matrix of order \(s\). Every self-stress the actuators can add is then \(\mathbf{N}\mathbf{w}\), where the vector \(\mathbf{w}\) holds one weight per state, and the force it adds to member \(k\) is the \(k\)-th row of \(\mathbf{N}\) applied to those weights,

\[
\Delta t_k = \mathbf{n}_k^{\mathsf{T}}\,\mathbf{w} \tag{3}
\]

\textbf{Reach.} The reach of member \(k\) is defined as the squared length
of that row,

\[
r_k = \lVert \mathbf{n}_k \rVert^{2} = \left(\mathbf{N}\mathbf{N}^{\mathsf{T}}\right)_{kk}, \qquad 0 \le r_k \le 1, \qquad \sum_{k=1}^{m} r_k = s \tag{4}
\]

The matrix \(\mathbf{N}\mathbf{N}^{\mathsf{T}}\) is the orthogonal projector onto the self-stress space, so \(r_k\) does not depend on which orthonormal basis is
chosen. It is the static indeterminacy factor of Eriksson and Tibert
{[}18{]}, which is the diagonal of the redundancy matrix introduced by
Ströbel {[}21{]} and developed by von Scheven and co-workers
{[}22,23{]}, evaluated with the same flexibility in every member so that
it depends on the geometry of the structure and not on the member sizes.
A member with \(r_k = 0\) has a row of zeros in \(\mathbf{N}\), so by Eq. (3) no
self-stress changes its force. The octet two-cell has \(m = 64\) and \(s = 10\), and twelve of its members have \(r_k = 0\). The basis must be orthonormal, since with any other basis of the same space the row lengths change with the choice of basis, whereas \(\mathbf{N}\mathbf{N}^{\mathsf{T}}\) does not.

\textbf{The self-stress test.} Let \(K\) be the set of members at the limit
in a load case which governs the best fixed design, where a member is
said to be at the limit when its utilisation equals the worst
utilisation of that design in that case, which is a property of the
design and not a statement that the member is at capacity, and let the utilisation of member \(k\) be its force divided by its capacity, which is the area \(a_k\) in tension, with the yield stress taken as one, and the buckling capacity \(c_k = \min(a_k,\, \kappa a_k^{2}/L_k^{2})\) in compression. Since the force added by a self-stress is linear in the weights \(\mathbf{w}\) of Eq. (3), so is the change in utilisation, and for member \(k\) it is

\[
\Delta u_k = \mathbf{g}_k^{\mathsf{T}}\,\mathbf{w}, \qquad
\mathbf{g}_k = \frac{\mathbf{n}_k}{a_k}\ \text{in tension}, \qquad
\mathbf{g}_k = -\frac{\mathbf{n}_k}{c_k}\ \text{in compression} \tag{5}
\]

where \(\mathbf{g}_k\) is the utilisation gradient of member \(k\), one entry per self-stress state, its branch set by the sign of the member's force in the fixed design, \(L_k\) is the member length and \(\kappa\) the buckling constant defined under Capacity below. The worst utilisation of the case falls only if one \(\mathbf{w}\) lowers every member of \(K\) at once, which is tested by the linear programme in the weights \(\mathbf{w}\) and a margin \(\delta\), the amount by which the member of \(K\) lowered least is lowered,

\[
\max_{\mathbf{w},\,\delta} \; \delta
\quad\text{subject to}\quad
\mathbf{g}_k^{\mathsf{T}}\mathbf{w} + \delta \le 0 \;\; (k \in K), \qquad
-1 \le w_i \le 1, \qquad \delta \ge 0 \tag{6}
\]

whose optimal value \(\delta^{\ast}\) is the margin attained. The bounds \(-1 \le w_i \le 1\) on the entries of \(\mathbf{w}\) only fix the scale, because any multiple of a self-stress is also a self-stress, and the bound \(\delta \ge 0\) excludes nothing since \(\mathbf{w} = \mathbf{0}\) is always feasible with \(\delta = 0\), so the sign of \(\delta^{\ast}\) is what matters. A positive margin gives a direction which lowers every member of \(K\), while \(\delta^{\ast} = 0\) means no such direction exists and the premium
is exactly zero. The two blocks of Figure 4 are the two ways this
happens, and by the theorem of the alternative \(\delta^{\ast} = 0\) holds if and only if multipliers \(\lambda_k \ge 0\), not all zero, exist with

\[
\sum_{k \in K} \lambda_k\, \mathbf{g}_k = \mathbf{0} \tag{7}
\]

with the gradients \(\mathbf{g}_k\) of Eq. (5), so the members at the limit pull against each other with no room left over {[}32,33{]}. A member of zero reach has \(\mathbf{g}_k = \mathbf{0}\) by Eq. (5), which
satisfies Eq. (7) on its own and is the first block, while the second
block needs at least two members with opposing gradients.

\textbf{Proposition (zero premium).} The geometry and the member sizes are fixed and the self-stress is chosen afresh for each load case. The utilisation of one member is a straight line in the self-stress weights \(\mathbf{w}\) while the member stays in tension and a steeper straight line while it stays in compression, with a corner where its force passes through zero. The worst utilisation of a load case is the highest of these lines over all its members at each \(\mathbf{w}\), which is a convex and piecewise linear function of \(\mathbf{w}\). The premium is exactly zero if and only if some load case which governs the fixed design admits multipliers \(\lambda_k \ge 0\), not all zero, which satisfy Eq. (7) over its members at the limit. \emph{Proof.} With the \(\lambda_k\) scaled to sum to one, Eq. (7) says that the gradients of the members at the limit balance each other and add to zero. For a convex function, that is the condition for \(\mathbf{w} = \mathbf{0}\) to be its lowest point, so no self-stress of any size lowers the worst utilisation of that case. The case stays at the worst value of the fixed design and holds the premium at zero. If no governing case admits Eq. (7), then by the theorem of the alternative each has a direction along which every member at its limit falls, and a small step along it lowers that case below the worst value of the fixed design. The self-stress is chosen for each case on its own and every other case already sits below that value, so the premium is positive.

The multipliers \(\boldsymbol{\lambda}\) of Eq. (7) form the certificate whose existence Figure 9 reports: they are found by a second linear programme, normalised to sum to one, and every certificate found leaves a residual below 10\(^{-15}\) of the largest gradient. Members with utilisation \(u \ge (1-10^{-6})\,u_{\max}\) of their case count as at the limit, and cases with \(u_{\max} \ge (1-10^{-9})\,U\) count as governing, where \(u_{\max}\) is the worst utilisation over the members of the case and \(U\) the worst over all members and all load cases. Equation (6)
is solved with the HiGHS simplex code through SciPy, and every zero
reported here was also obtained independently from the premium itself,
which is computed by Eq. (8) with the member sizes frozen.

\textbf{Load set.} Every structure is loaded at one node at a time by a
unit force in each of twenty directions in turn and each direction is
one load case, so a design is sized against twenty cases and the premium
is taken over the same twenty. The directions are listed in
Supplementary Table S1. For the three-dimensional structures they are
ten points spread over the upper hemisphere together with their exact
reverses and for the planar structures they are spaced evenly round the
circle, and the planar set is also closed under both mirrors of the plane. The set was fixed before any result reported here was computed and every number in the paper is on it. Every direction in the set comes with its exact opposite, so no sized design can lean one way because of the loads. The count of twenty sets how finely the directions are covered and the proposition holds for any finite set of load cases. The eight structures on this set are therefore worked examples, and the test applies in the same way to any other set. A zero from a member of zero reach holds wherever this member stays at the limit in the governing case, while the coupled zeros and every positive premium reported here are values on this set. In the sweep of Figure
7 each free node of each structure is loaded in turn in the same twenty
directions and the fixed design is sized afresh for that node.

\textbf{Best fixed design.} The fixed competitor of every comparison is
sized against the whole load set at the material budget. The budget is
the sum of area times length over the members and every member keeps a lower limit on its area of one part in ten thousand of the uniform area, so that no member is removed by the sizing. The sizing is a stress-ratio iteration in which each pass finds the worst utilisation \(u_k\) of every member over the load cases and the worst \(U\) of these over the design, which is the largest utilisation over all members and all load cases and is the single number by which every design in this paper is judged, multiplies each area by \(\sqrt{u_k/U}\) so that under-worked members shrink while the governing member keeps its area, and then scales all areas by one factor to return to the budget. This is the fully-stressed design update with the ratio damped by its square root, which slows the step and keeps the iteration from oscillating. The passes repeat until the worst utilisation changes by less than one part in 10\(^{12}\) between passes, and the whole run is relaunched from its end point up to twelve times. The iteration is a local search and where it ends depends on where it starts, so it is run from two starting points and the better result is kept. One start is the uniform design. The other is the solution of a linear programme which relaxes the sizing problem in two ways, by replacing the buckling capacity, which is curved in the area, with the straight-line function sitting just above it, and by letting the member forces take any values that balance the load without the compatibility the structure imposes. Both relaxations can only lower the worst utilisation, so the value of that programme is a lower bound on what any design at the budget can attain with or without actuators. That bound sits 17\% to 84\% below the fixed
design across the eight structures of Figure 7, so it is loose and does
not certify the fixed design as optimal. The search was therefore tested
in two further ways on the eight structures of Figure 7, each at its
loaded node of Figure 9. Twenty further starts, sixteen drawn at random over
the material budget and four perturbed from the design found, all
returned that design to within two parts in 10\(^{14}\) and with the
same premium. An independent gradient optimiser, the method of moving
asymptotes on a \(p\)-norm of the member utilisations, could not improve on
that design when started from it and ended between 0.02\% and 16\% worse
when started from the uniform design. No better fixed design was found,
which is evidence and not a proof that none exists. The adaptive side
needs no search because the best self-stress for a load case is an exact
linear programme over the whole self-stress space. A better fixed design
would change every positive premium, and on the three structures of
Figure 2 the move from the uniform design to the best fixed design
lowered it in each case, while each zero is proved at the fixed design
in use by the certificate of Eq. (7). The self-stress test
itself holds for any fixed design, so a better design could change which
members stand at the limit and would leave the test as it is.

\textbf{Best actuation.} The adaptive design keeps the member sizes of
the best fixed design and adds one self-stress per load case, chosen
afresh for each case and with no limit on the number of actuators or on
the force they apply, which is the retrofit comparison set out in the
Results. For a load case with fixed-design member forces \(\mathbf{t}\) the best actuation is the \(\mathbf{w}\) which gives the lowest worst utilisation over the members of that case,

\[
\min_{\mathbf{w},\,U} \; U
\quad\text{subject to}\quad
(\mathbf{t} + \mathbf{N}\mathbf{w})_k \le U a_k, \qquad -(\mathbf{t} + \mathbf{N}\mathbf{w})_k \le U c_k \qquad (k = 1, \dots, m) \tag{8}
\]

where \(U\) is the same worst utilisation taken over the members for one load case once the actuators act, and its lowest value \(U^{\ast}\) and the \(\mathbf{w}^{\ast}\) which attains it are the best actuation, with the areas held fixed.
The premium is the fraction by which the worst of these values over the
whole load set falls below the worst utilisation of the best fixed
design. Where several combinations attain that same lowest value the
choice is not unique, as it is on the octet, and the one returned by the optimiser is what Figure 3 draws.

\textbf{Load factor.} Member forces are linear in the applied load and
member capacities do not depend on it, so multiplying the load by a
factor multiplies every utilisation by that factor and a design stays
within capacity exactly while the factor is at most the reciprocal of
its worst utilisation. The load factor of a design is that limit,

\[
\Lambda = 1/U, \qquad R = 1 - \Lambda_{\mathrm{fix}}/\Lambda_{\mathrm{ad}} \tag{9}
\]

where \(U\) is the worst utilisation of the design over the load set and \(R\) the premium. The load factor is quoted against a reference load, taken as the largest load the best fixed design of that structure carries, so that design reads 1.00 and a design which reads 1.27 carries 1.27 times as much before its
worst member arrives at capacity. Nothing in the comparison changes
under this reporting, because rescaling the load multiplies the worst
utilisation of the fixed design and of the actuated design alike and the
premium is their ratio. The proportionality is exact here since the
analysis is linear elastic and the capacities are set by area and length
alone, and it would not hold if plasticity or large displacements
entered.

\textbf{Lifetime comparison.} The adaptive design can apply zero
self-stress in any load case, so with the member areas of the fixed
design its worst utilisation is no greater than that of the fixed design
in every load case and for every load direction, not only the twenty of
the load set. Under a strength limit the service life ends at the first
load event whose utilisation attains one. The failing events of the
adaptive design are then a subset of those of the fixed design, so on every sequence of load events the adaptive design fails no earlier whatever the dependence between the events. This assumes that the actuators apply the best self-stress for each load event before any member is loaded to its capacity. That statement needs no load
model, and where the events are independent with exceedance probability \(q\) the survival probability over \(n\) events is \((1-q)^{n}\) and the expected life is \(1/q\), and a lower worst utilisation lowers \(q\) under any
load law. The dependence between real load events is unknown and the
independent form is not used anywhere in the paper. The account assumes
that a survived event leaves the structure as built, which excludes
cumulative damage, and it compares the two designs at the same member
areas, which is the retrofit comparison.

\textbf{Capacity and slenderness.} Each member is a pin-ended bar with a
unit yield stress, so its capacity in tension is its area and its
capacity in compression is the smaller of its area and its Euler load.
The Euler load of member \(k\) is \(\kappa a_k^{2}/L_k^{2}\), which is the
Euler law for a solid circular section written in the units of the
model, and \(\kappa\) is the one constant the model does not derive. It is set
for each structure so that the median member of the uniform design has a chosen ratio \(\rho\) of Euler load to yield load, and every other member then
has its own ratio in proportion to \(a_k/L_k^{2}\). For steel with \(E = 210\) GPa and \(\sigma_y = 355\) MPa a ratio \(\rho\) corresponds to a slenderness ratio \(L/r_g = (\pi^{2}E/\rho\sigma_y)^{1/2} = (5838/\rho)^{1/2}\), where \(r_g\) is the radius of gyration of the section, which is the axis of Figure 6, and \(\rho = 0.7\) (\(L/r_g = 91\)) is used in Figures 2 to 5. The
utilisation of a member is the force it carries divided by its capacity,
so it exceeds one where the member has failed, and the worst utilisation
over every member and every load case is what the fixed design is sized
to minimise and what the actuators lower.

\textbf{Finite-element check.} The hub of Figure 6 at slenderness 653 was built in Abaqus as twelve two-node truss elements with \(E = 210\) GPa and \(\sigma_y = 355\) MPa, the apex 1000 mm above an anchor circle of radius 1000
mm so that every member is 1414 mm long, and each member was given the
area of the best fixed design from the reduced model. A load of 1000 N
was applied at the apex along each of the twenty directions of the
paper's load set in turn, which gives the fixed design in twenty load
steps, and the actuated design was run as twenty further analyses in
which the optimal self-stress of the reduced model for that direction
was imposed as an initial stress in each member before the load was
applied. Member forces were read from Abaqus and converted to
utilisations with the same capacity law, the yield stress in tension and
the smaller of the yield stress and the Euler stress \(\pi E A/(4L^{2})\) of a
pin-ended solid circular bar in compression, and the premium was formed
from the worst utilisations exactly as in the reduced model. Each input file was first solved directly, since the hub with one free node is three linear equations, and the check was set to fail if Abaqus differed from that solution by more than 0.5\%.

\textbf{Use of AI tools.} AI tools (Claude, Anthropic) were used in developing the analysis code and in preparing the manuscript. The author designed the study, directed and checked all computations, and takes full responsibility for the content.

\hypertarget{data-and-code-availability}{%
\subsection{Data and code
availability}\label{data-and-code-availability}}

The reduced model is written in Python with NumPy and SciPy and every
linear programme is solved with the HiGHS code through SciPy. The
scripts which build each structure and size the fixed design and solve
the actuation and certificate programmes and draw every figure will be
deposited in a public repository with a digital object identifier on publication, together with the result files they wrote and the Abaqus
input files of the finite-element check. Each result file carries a stamp which names its metric, its aggregation, its comparator, its intervention, its optimiser and its load set. In the code the certificate programme of Eq. (7) works on the signed rows of \(\mathbf{N}\) rather than on \(\mathbf{g}_k\), which scales each multiplier by the positive capacity of its member and leaves existence unchanged.

\hypertarget{acknowledgements}{%
\subsection{Acknowledgements}\label{acknowledgements}}

The author received no specific funding for this work.

\hypertarget{author-contributions}{%
\subsection{Author contributions}\label{author-contributions}}

I.M. conceived the study, ran the computations, checked the results and wrote the paper.

\hypertarget{competing-interests}{%
\subsection{Competing interests}\label{competing-interests}}

The author declares no competing interests.

\hypertarget{references}{%
\subsection{References}\label{references}}

\begin{enumerate}
\def\labelenumi{\arabic{enumi}.}
\tightlist
\item
  Florijn, B., Coulais, C. \& van Hecke, M. Programmable mechanical
  metamaterials. \emph{Phys. Rev.~Lett.} \textbf{113}, 175503 (2014).
\item
  Rocks, J. W., Ronellenfitsch, H., Liu, A. J., Nagel, S. R. \&
  Katifori, E. Limits of multifunctionality in tunable networks.
  \emph{Proc. Natl Acad. Sci. USA} \textbf{116}, 2506--2511 (2019).
\item
  Bossart, A., Dykstra, D. M. J., van der Laan, J. \& Coulais, C.
  Oligomodal metamaterials with multifunctional mechanics. \emph{Proc.
  Natl Acad. Sci. USA} \textbf{118}, e2018610118 (2021).
\item
  Bordiga, G. et al.~Automated discovery of reprogrammable nonlinear
  dynamic metamaterials. \emph{Nat. Mater.} \textbf{23}, 1486--1494
  (2024).
\item
  Du, Y., van Mastrigt, R., Veenstra, J. \& Coulais, C. Metamaterials
  that learn to change shape. \emph{Nat. Phys.} \textbf{22}, 784--790
  (2026).
\item
  Shaikeea, A. J. D., Cui, H., O'Masta, M., Zheng, X. R. \& Deshpande,
  V. S. The toughness of mechanical metamaterials. \emph{Nat. Mater.}
  \textbf{21}, 297--304 (2022).
\item
  Pham, M.-S., Liu, C., Todd, I. \& Lertthanasarn, J. Damage-tolerant
  architected materials inspired by crystal microstructure.
  \emph{Nature} \textbf{565}, 305--311 (2019).
\item
  Gu, J. et al.~Optimization and control of actuator networks in
  variable geometry truss systems using genetic algorithms. \emph{Nat.
  Commun.} \textbf{16}, 8432 (2025).
\item
  Kirsch, U. \& Moses, F. Optimization of structures with control forces
  and displacements. \emph{Eng. Optim.} \textbf{3}, 37--44 (1977).
\item
  Senatore, G. Designing and prototyping adaptive structures: an
  energy-based approach beyond lightweight design. In \emph{Robotic
  Building} (ed.~Bier, H.) 169--189 (Springer, 2018).
\item
  Wang, Y. \& Senatore, G. Minimum energy adaptive structures:
  all-in-one problem formulation. \emph{Comput. Struct.} \textbf{236},
  106266 (2020).
\item
  Reksowardojo, A. P. \& Senatore, G. Design of ultra-lightweight and
  energy-efficient civil structures through shape morphing.
  \emph{Comput. Struct.} \textbf{289}, 107149 (2023).
\item
  Senatore, G. \& Wang, Y. Topology optimization of adaptive structures:
  new limits of material economy. \emph{Comput. Methods Appl. Mech.
  Eng.} \textbf{422}, 116710 (2024).
\item
  Adam, B. \& Smith, I. F. C. Self-diagnosis and self-repair of an
  active tensegrity structure. \emph{J. Struct. Eng.} \textbf{133},
  1752--1761 (2007).
\item
  Senatore, G., Duffour, P. \& Winslow, P. Synthesis of minimum energy
  adaptive structures. \emph{Struct. Multidiscip. Optim.} \textbf{60},
  849--877 (2019).
\item
  Steffen, S., Weidner, S., Blandini, L. \& Sobek, W. Using influence
  matrices as a design and analysis tool for adaptive truss and beam
  structures. \emph{Front. Built Environ.} \textbf{6}, 83 (2020).
\item
  Steffen, S., Blandini, L. \& Sobek, W. Analysis of the inherent
  adaptability of basic truss and frame modules by means of an extended
  method of influence matrices. \emph{Eng. Struct.} \textbf{266}, 114588
  (2022).
\item
  Eriksson, A. \& Tibert, A. G. Redundant and force-differentiated
  systems in engineering and nature. \emph{Comput. Methods Appl. Mech.
  Eng.} \textbf{195}, 5437--5453 (2006).
\item
  Wicks, B. \& Guest, S. D. Single member actuation in large repetitive
  truss structures. \emph{Int. J. Solids Struct.} \textbf{41}, 965--978 (2004).
\item
  Symons, D. D., Hutchinson, R. G. \& Fleck, N. A. Actuation of the
  Kagome double-layer grid. Part 1: Prediction of performance of the
  perfect structure. \emph{J. Mech. Phys. Solids} \textbf{53}, 1855--1874 (2005).
\item
  Ströbel, D. \emph{Die Anwendung der Ausgleichungsrechnung auf
  elastomechanische Systeme}. Dissertation, Deutsche Geodätische
  Kommission, Reihe C, Heft 478 (München, 1997).
\item
  von Scheven, M., Ramm, E. \& Bischoff, M. Quantification of the
  redundancy distribution in truss and beam structures. \emph{Int. J.
  Solids Struct.} \textbf{213}, 41--49 (2021).
\item
  Forster, D. \& von Scheven, M. The redundancy matrix as a performance
  indicator for structural assessment. \emph{J. Theor. Comput. Appl.
  Mech.} https://doi.org/10.46298/jtcam.13574 (2025).
\item
  Senatore, G. \& Reksowardojo, A. P. Force and shape control strategies
  for minimum energy adaptive structures. \emph{Front. Built Environ.}
  \textbf{6}, 105 (2020).
\item
  Melchers, R. E. \& Beck, A. T. \emph{Structural Reliability Analysis and
  Prediction} 3rd edn, Ch. 6 (Wiley, 2018).
\item
  Senatore, G., Virgili, A. \& Blandini, L. Global optimal actuator
  placement for adaptive structures: new formulation and benchmarking.
  \emph{J. Intell. Mater. Syst. Struct.} \textbf{36}, 151--173 (2025).
\item
  Efinger, D., Ostertag, A., Dazer, M., Borschewski, D., Albrecht, S. \&
  Bertsche, B. Reliability as a key driver for a sustainable design of
  adaptive load-bearing structures. \emph{Sustainability} \textbf{14}, 895 (2022).
\item
  Efinger, D., Canny, D., Dazer, M., Blandini, L. \& Senatore, G.
  High-fidelity analysis of service life extension for steel beam
  bridges retrofitted with external adaptive tensioning systems. \emph{Struct.
  Infrastruct. Eng.} https://doi.org/10.1080/15732479.2026.2631164
  (2026).
\item
  Calladine, C. R. Buckminster Fuller's ``tensegrity'' structures and
  Clerk Maxwell's rules for the construction of stiff frames. \emph{Int.
  J. Solids Struct.} \textbf{14}, 161--172 (1978).
\item
  Pellegrino, S. \& Calladine, C. R. Matrix analysis of statically and
  kinematically indeterminate frameworks. \emph{Int. J. Solids Struct.}
  \textbf{22}, 409--428 (1986).
\item
  Pellegrino, S. Structural computations with the singular value
  decomposition of the equilibrium matrix. \emph{Int. J. Solids Struct.}
  \textbf{30}, 3025--3035 (1993).
\item
  Gordan, P. Über die Auflösung linearer Gleichungen mit reellen
  Coefficienten. \emph{Math. Ann.} \textbf{6}, 23--28 (1873).
\item
  Schrijver, A. \emph{Theory of Linear and Integer Programming} (Wiley,
  1986).
\end{enumerate}

\newpage

\hypertarget{supplementary-information}{%
\section{Supplementary Information}\label{supplementary-information}}

\textbf{Supplementary Table S1.} The twenty load directions used throughout the
paper. (a) For the three-dimensional structures, unit vectors spread
over the sphere with ten on the upper hemisphere and their ten exact
reverses, so that every direction has its opposite in the set, ten
places apart in the numbering. Azimuth is measured from the x-axis in
the horizontal plane and elevation from that plane. (b) For the planar
structures, twenty directions spaced evenly round the circle from the
x-axis, with the same pairing. Every direction is applied at the loaded
node as a unit load in turn, and each direction is one load case.

\begin{enumerate}
\def\labelenumi{(\alph{enumi})}
\tightlist
\item
  Three-dimensional structures
\end{enumerate}

\begin{longtable}[]{@{}lllllll@{}}
\toprule\noalign{}
Direction & x & y & z & Azimuth (°) & Elevation (°) & Reverse \\
\midrule\noalign{}
\endhead
\bottomrule\noalign{}
\endlastfoot
1 & 0.3122 & 0.0000 & 0.9500 & 0.0 & 71.8 & 11 \\
2 & $-$0.3884 & 0.3558 & 0.8500 & 137.5 & 58.2 & 12 \\
3 & 0.0578 & $-$0.6589 & 0.7500 & 275.0 & 48.6 & 13 \\
4 & 0.4624 & 0.6031 & 0.6500 & 52.5 & 40.5 & 14 \\
5 & $-$0.8224 & $-$0.1455 & 0.5500 & 190.0 & 33.4 & 15 \\
6 & 0.7535 & $-$0.4793 & 0.4500 & 327.5 & 26.7 & 16 \\
7 & $-$0.2432 & 0.9046 & 0.3500 & 105.0 & 20.5 & 17 \\
8 & $-$0.4463 & $-$0.8593 & 0.2500 & 242.6 & 14.5 & 18 \\
9 & 0.9287 & 0.3392 & 0.1500 & 20.1 & 8.6 & 19 \\
10 & $-$0.9232 & 0.3811 & 0.0500 & 157.6 & 2.9 & 20 \\
11 & $-$0.3122 & 0.0000 & $-$0.9500 & 180.0 & $-$71.8 & 1 \\
12 & 0.3884 & $-$0.3558 & $-$0.8500 & 317.5 & $-$58.2 & 2 \\
13 & $-$0.0578 & 0.6589 & $-$0.7500 & 95.0 & $-$48.6 & 3 \\
14 & $-$0.4624 & $-$0.6031 & $-$0.6500 & 232.5 & $-$40.5 & 4 \\
15 & 0.8224 & 0.1455 & $-$0.5500 & 10.0 & $-$33.4 & 5 \\
16 & $-$0.7535 & 0.4793 & $-$0.4500 & 147.5 & $-$26.7 & 6 \\
17 & 0.2432 & $-$0.9046 & $-$0.3500 & 285.0 & $-$20.5 & 7 \\
18 & 0.4463 & 0.8593 & $-$0.2500 & 62.6 & $-$14.5 & 8 \\
19 & $-$0.9287 & $-$0.3392 & $-$0.1500 & 200.1 & $-$8.6 & 9 \\
20 & 0.9232 & $-$0.3811 & $-$0.0500 & 337.6 & $-$2.9 & 10 \\
\end{longtable}

\begin{enumerate}
\def\labelenumi{(\alph{enumi})}
\setcounter{enumi}{1}
\tightlist
\item
  Planar structures
\end{enumerate}

\begin{longtable}[]{@{}lllll@{}}
\toprule\noalign{}
Direction & Angle from the x-axis (°) & x & y & Reverse \\
\midrule\noalign{}
\endhead
\bottomrule\noalign{}
\endlastfoot
1 & 0 & 1.0000 & 0.0000 & 11 \\
2 & 18 & 0.9511 & 0.3090 & 12 \\
3 & 36 & 0.8090 & 0.5878 & 13 \\
4 & 54 & 0.5878 & 0.8090 & 14 \\
5 & 72 & 0.3090 & 0.9511 & 15 \\
6 & 90 & 0.0000 & 1.0000 & 16 \\
7 & 108 & $-$0.3090 & 0.9511 & 17 \\
8 & 126 & $-$0.5878 & 0.8090 & 18 \\
9 & 144 & $-$0.8090 & 0.5878 & 19 \\
10 & 162 & $-$0.9511 & 0.3090 & 20 \\
11 & 180 & $-$1.0000 & 0.0000 & 1 \\
12 & 198 & $-$0.9511 & $-$0.3090 & 2 \\
13 & 216 & $-$0.8090 & $-$0.5878 & 3 \\
14 & 234 & $-$0.5878 & $-$0.8090 & 4 \\
15 & 252 & $-$0.3090 & $-$0.9511 & 5 \\
16 & 270 & 0.0000 & $-$1.0000 & 6 \\
17 & 288 & 0.3090 & $-$0.9511 & 7 \\
18 & 306 & 0.5878 & $-$0.8090 & 8 \\
19 & 324 & 0.8090 & $-$0.5878 & 9 \\
20 & 342 & 0.9511 & $-$0.3090 & 10 \\
\end{longtable}

\textbf{Supplementary note on the load set.} The ten upper-hemisphere directions
of Table S1(a) are the first ten of twenty points placed on the sphere
by the golden-angle rule, in which point \(i\) of \(m\) has height \(z = 1 - 2(i + \tfrac12)/m\) and azimuth \(i\) times the golden angle \(\pi(3 - \sqrt{5})\), which spreads the
points evenly in area. The ten with z above zero are kept and their
reverses added, which gives a set closed under reversal without the
reflection symmetry a regular polyhedron would impose. The planar set of
Table S1(b) is closed under both mirrors of the plane, which an odd
count cannot be, and the count of twenty was fixed before any result in
the paper was computed.

\includegraphics[width=1\textwidth,height=\textheight]{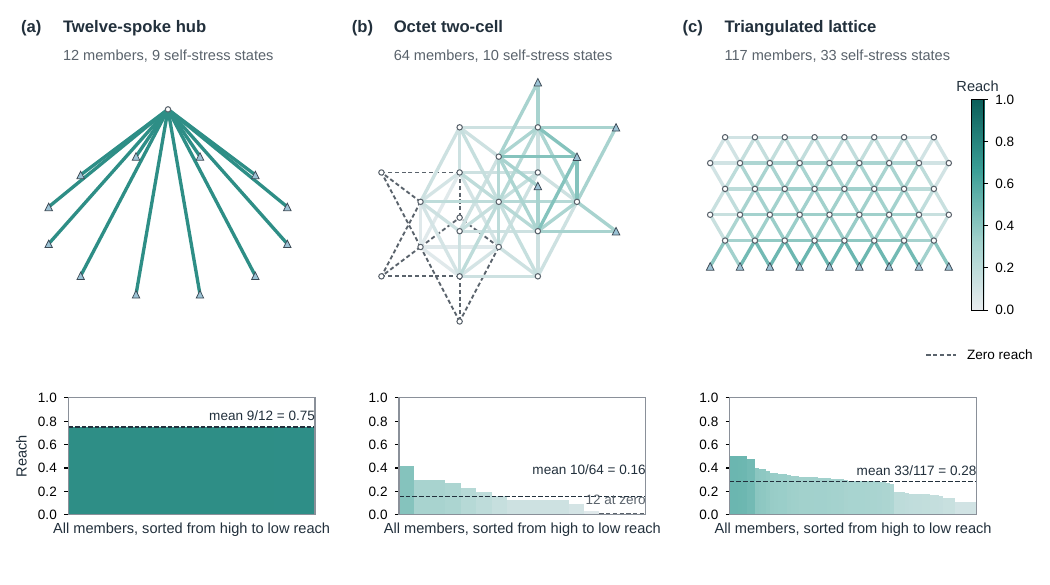}

\textbf{Supplementary Figure S1.} Reach of every member in the three
structures of Figure 3. Reach is the diagonal entry \(r_k = (\mathbf{N}\mathbf{N}^{\mathsf{T}})_{kk}\) of the projector onto the self-stress space
(Methods, Eq. (4)), and it depends on the geometry alone, so no loads or
member sizes enter this figure. Reach is therefore the same whichever
node is loaded and in whichever direction, and one set of values serves
every panel of Figure 4 and Supplementary Figure S7. It does change if the
supports change, since fixing a node removes its equilibrium equations from \(\mathbf{B}\). The loads decide only which members are at the limit, and so
whether a member of zero reach is among them. Top, each member is
coloured by its reach on one scale, with members of zero reach dashed.
Bottom, one bar per member, with the members sorted from highest to
lowest reach, and the dashed line marks the mean, which equals the
number of self-stress states divided by the number of members. (a) On
the twelve-spoke hub symmetry gives every spoke the same reach of 0.75.
(b) On the octet two-cell reach runs from 0.42 down to zero, and the
twelve members of zero reach all sit at the free end away from the
supports. (c) On the triangulated lattice every member can be reached,
with values from 0.11 to 0.51. A high reach does not by itself give a
premium, since the test of Figure 4 also asks whether the members at the
limit can be lowered together.

\includegraphics[width=1\textwidth,height=\textheight]{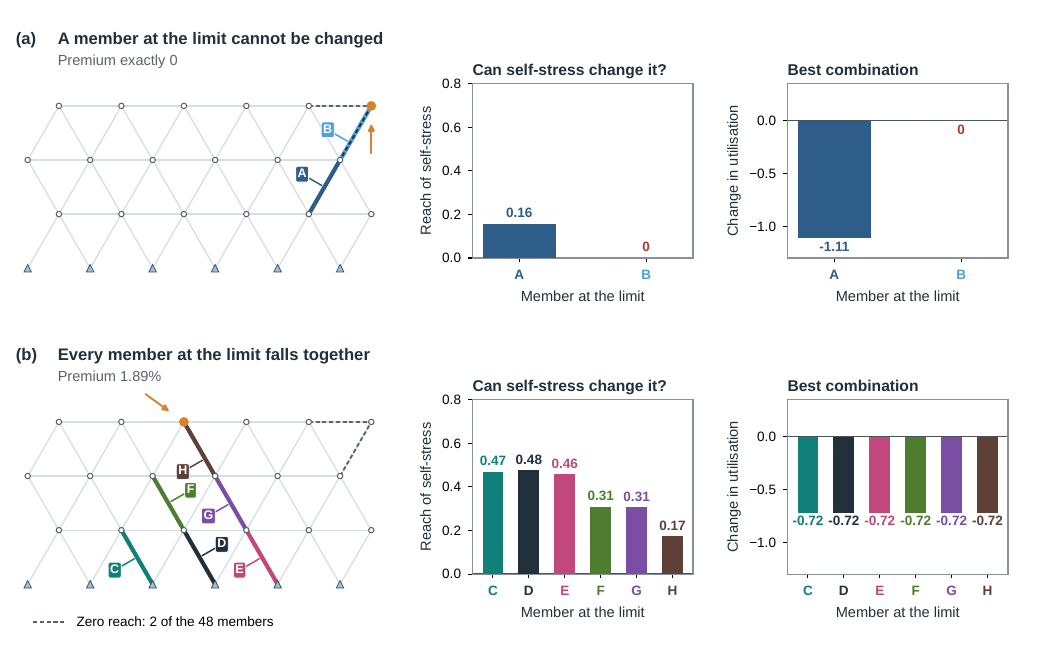}

\textbf{Supplementary Figure S2.} The same two blocks on a second
structure, a staggered grid of 48 members with 12 self-stress states.
Each row loads one node alone in twenty directions and draws the load
case which governs the best fixed design, with the loaded node in orange
and the members at the limit lettered and coloured as on the axes. (a)
Two members are at the limit and member B has zero reach, so no
combination of self-stress states changes its force and the worst
utilisation stays where B sets it, which holds the load factor at 1.00
and makes the premium exactly zero even though member A can be lowered
by more than the worst value. Two of the 48 members have zero reach,
drawn dashed, and both lie at the free corner of the grid. (b) At
another node six members are at the limit, all of them can be reached,
and the best combination lowers all six by the same amount, so the
premium is positive at 1.89\% and the load factor rises to 1.019. Reach
is the diagonal entry of the projector onto the self-stress space
(Methods, Eq. (4)) and depends on the geometry of the structure, where the
nodes sit as well as how the members join them and where the supports
are. The right-hand bars give the change in utilisation of each member
at the limit under the combination which solves Eq. (6), as a share of the worst value. Only the signs and ratios of the bars matter, so the size of a bar is not the premium. The premium is set by how far these members can fall before a member below the limit rises to meet them. The node is sized against its own best fixed design at the matched material budget, with Euler member capacity at median slenderness 91 and the member sizes frozen under actuation.

\includegraphics[width=1\textwidth,height=\textheight]{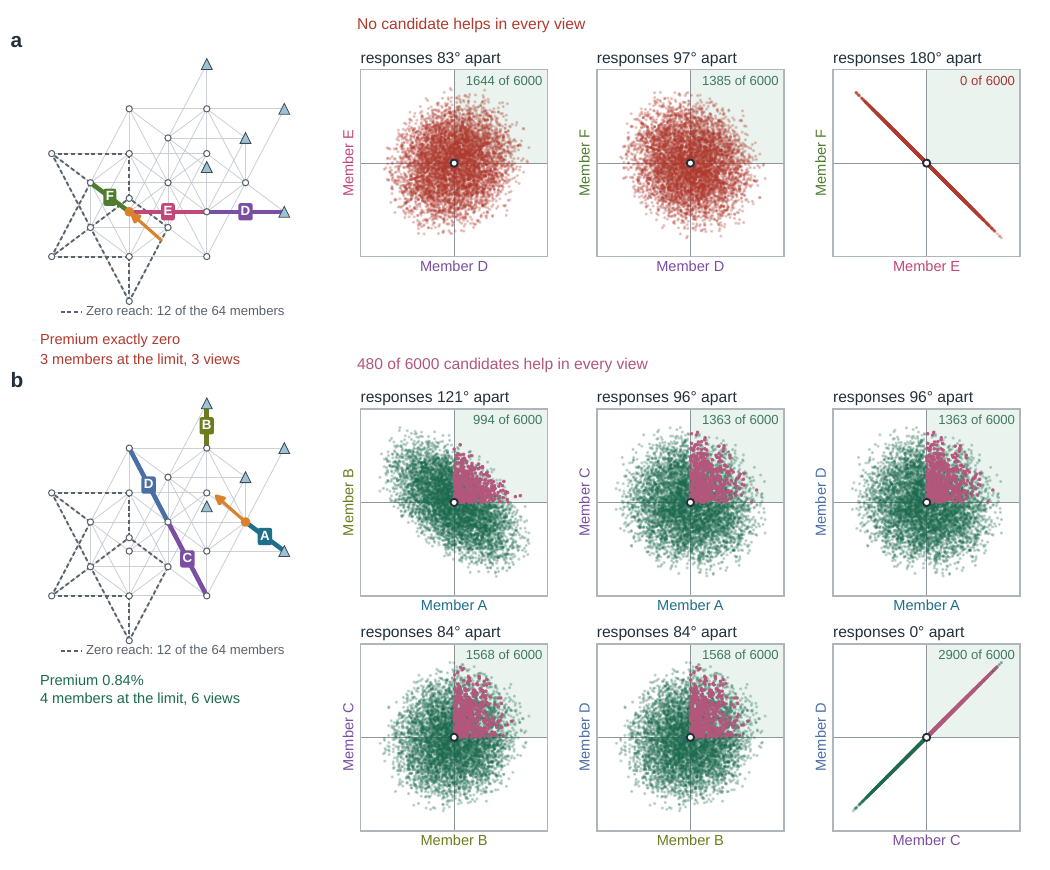}

\textbf{Supplementary Figure S3.} Every two-member view of the candidate
self-stresses on the octet, at a node where the premium vanishes (a) and at a node with a
positive premium (b). Six thousand candidate self-stresses are drawn
evenly over all mixtures of the ten states, and each panel plots the
change they make to the utilisation of two of the members at the limit,
scaled by its own range, so up and to the right means both members
improve and the shaded quarter is where a candidate helps both. The
drawing at the left shows the loaded node in orange with the governing
one of its twenty load directions as an arrow, the supports as triangles
and the members at the limit lettered and coloured as on the axes, in the view of Figure 4, with the twelve members of zero reach dashed as there. The angle above each panel is the angle between the responses of the two members to the candidates, not the angle between the members. (a) The node and load case of Figure 4c, with the same letters. Three members are at the limit and all three can be reached, but E and F change in exact opposition, so their view
collapses to a line which misses the shaded quarter and no candidate
helps all three. (b) Four members are at the limit and 480 of the 6000
candidates lie in the shaded quarter of every view, drawn in rose, so
the premium is positive at 0.84\%. The node is sized against its own best fixed design at the matched material budget, with Euler member capacity at slenderness 91 and the member sizes frozen under actuation.

\includegraphics[width=1\textwidth,height=\textheight]{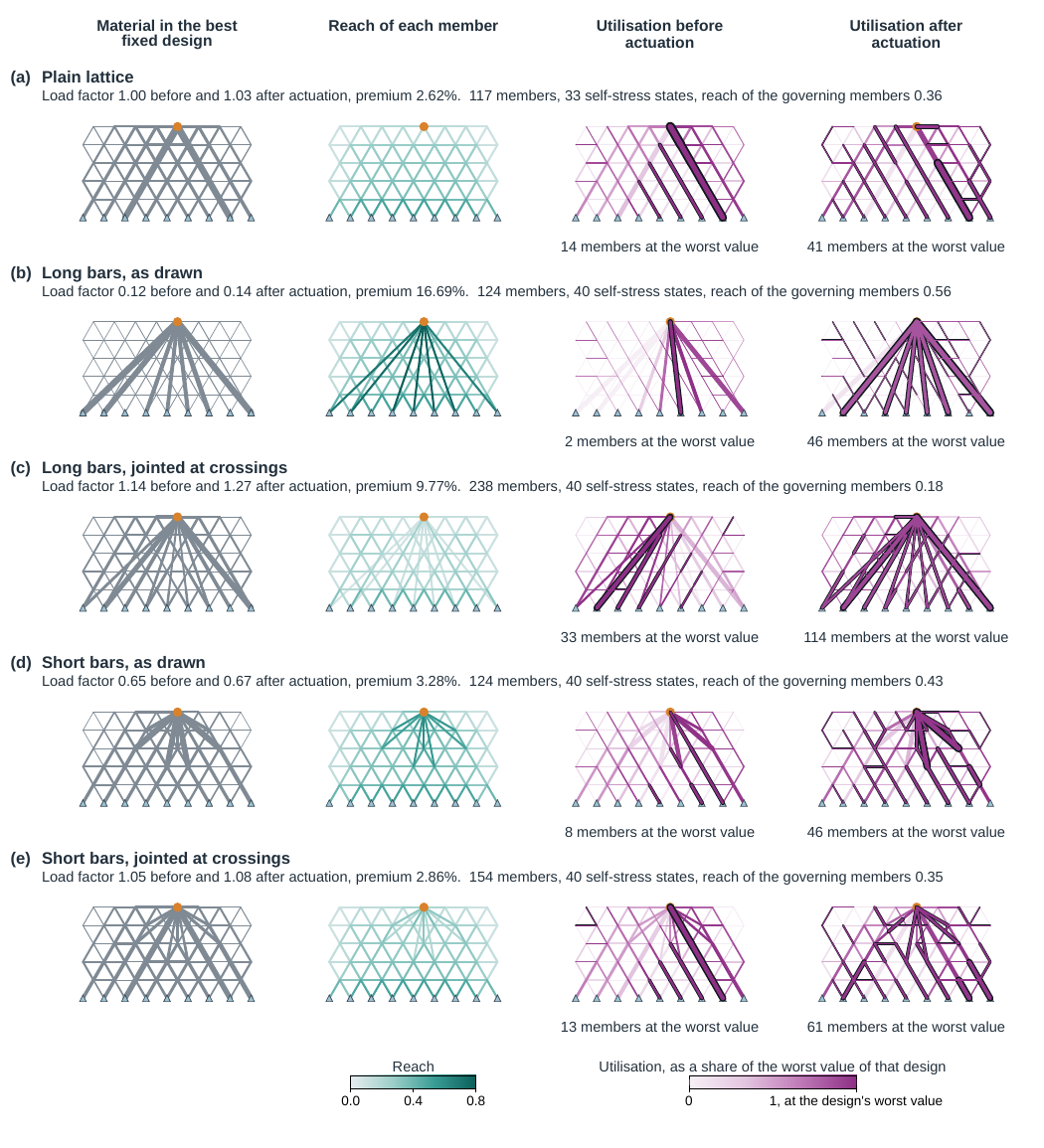}

\textbf{Supplementary Figure S4.} Where the material goes, how far the
self-stress reaches, and which members govern, for the five
configurations of Figure 5. Each row is one configuration, drawn at the load case which governs after actuation, which is the convention of Figure 3, and utilisation
is shown as a share of that design's worst value, so the colour is
comparable within a row rather than between rows. Column 1 gives the
member area of the best fixed design as line width, column 2 the reach
of every member on one scale, and columns 3 and 4 the member utilisation
before and after the best actuation, with the members at that design's
worst value outlined in black. The line above each row gives the load
factor before and after actuation against the reference load of Figure
5, the premium, the member and self-stress counts, and the mean reach of
the members which govern the fixed design. The mean reach of a whole
structure equals the number of self-stress states divided by the number
of members, so a jointed structure is paler by construction and only the
values on the governing members carry weight. The configuration with the
highest reach and the largest premium of the five, (b), is also the
weakest design, which is why reach is a condition for a premium and not
something to maximise. The set of members at the worst value grows under
actuation because the best actuation is a minimax which stops only when
no further lowering is possible, and in (c) that set is a connected path
from the loaded node to the supports. Euler member capacity, matched
material budget, buckling constant 81.9 throughout, twenty load
directions.

\includegraphics[width=1\textwidth,height=\textheight]{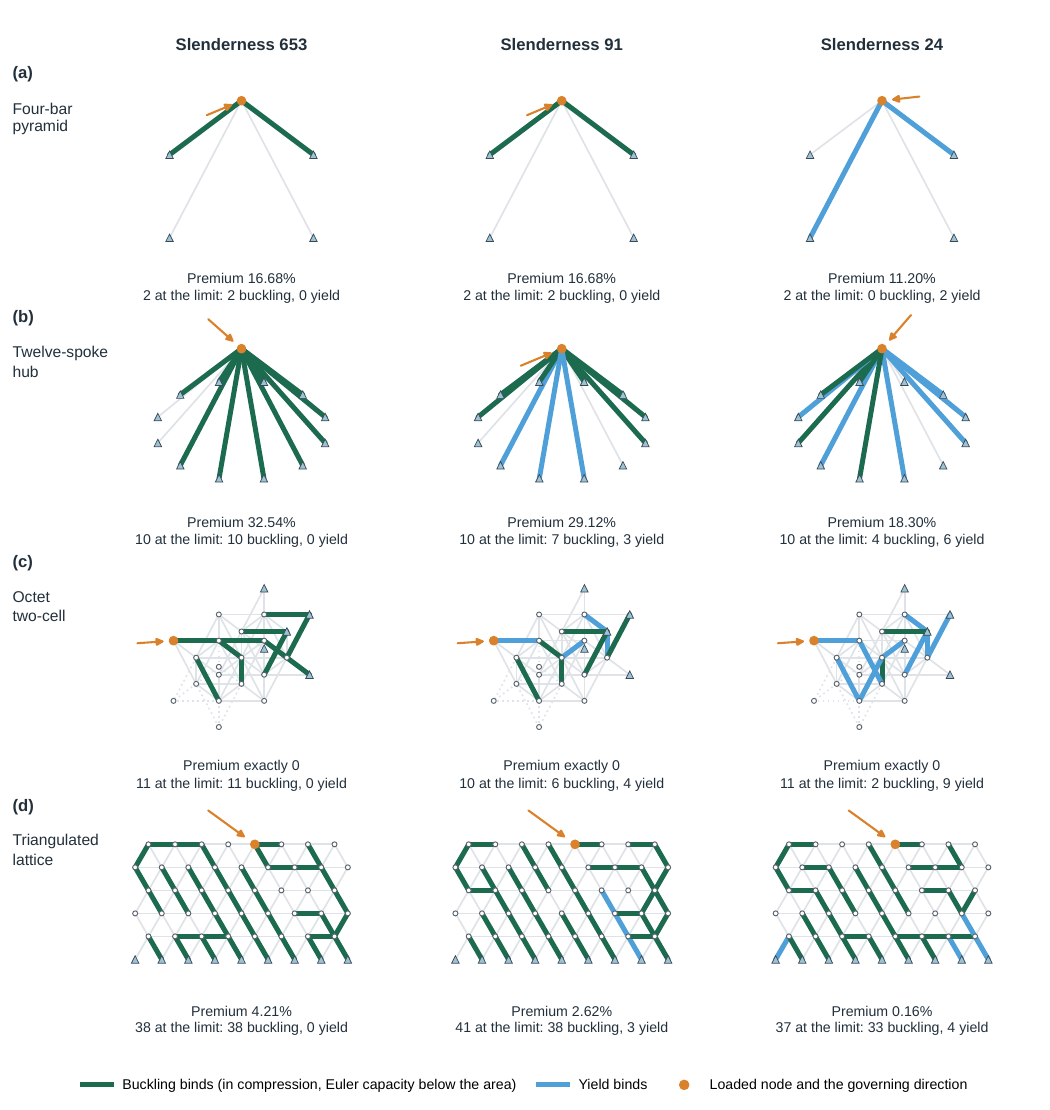}

\textbf{Supplementary Figure S5.} Which limit binds the members at the limit, at three slendernesses, for the four structures of Figure 1. Each
panel draws the load case which governs at that slenderness, with the
loaded node in orange and its governing direction as an arrow, and
colours only the members at the limit of the actuated design, green where
the member is in compression and its Euler capacity is below its area so
that buckling binds, and blue where yield binds. Under each panel is the
premium at that slenderness and the count of members at the limit by the
limit which binds them. (a) The pyramid has two members at its limit and
both change over from buckling to yield between slenderness 91 and 24,
which is the step in its curve in Figure 6. (b) The hub keeps ten
members at its limit throughout, all of them buckling-limited at 653,
seven at 91 and four at 24, so its changeover is spread over the sweep
and its premium slides. (c) The octet changes over in the same way and
its premium stays at exactly zero, since its block is one of reach
rather than of capacity, and the same load case governs at all three
slendernesses so its three panels are drawn at the one loaded node and
the one direction of Figures 3b and 4b.
(d) The lattice behaves like the hub over its 117 members. At
slenderness 653 every member at the limit of every design is
buckling-limited, which is the plateau of Figure 6. At every slenderness the buckling constant is calibrated afresh and the best fixed design is found afresh at the matched material budget, with the member sizes frozen under actuation.

\includegraphics[width=1\textwidth,height=\textheight]{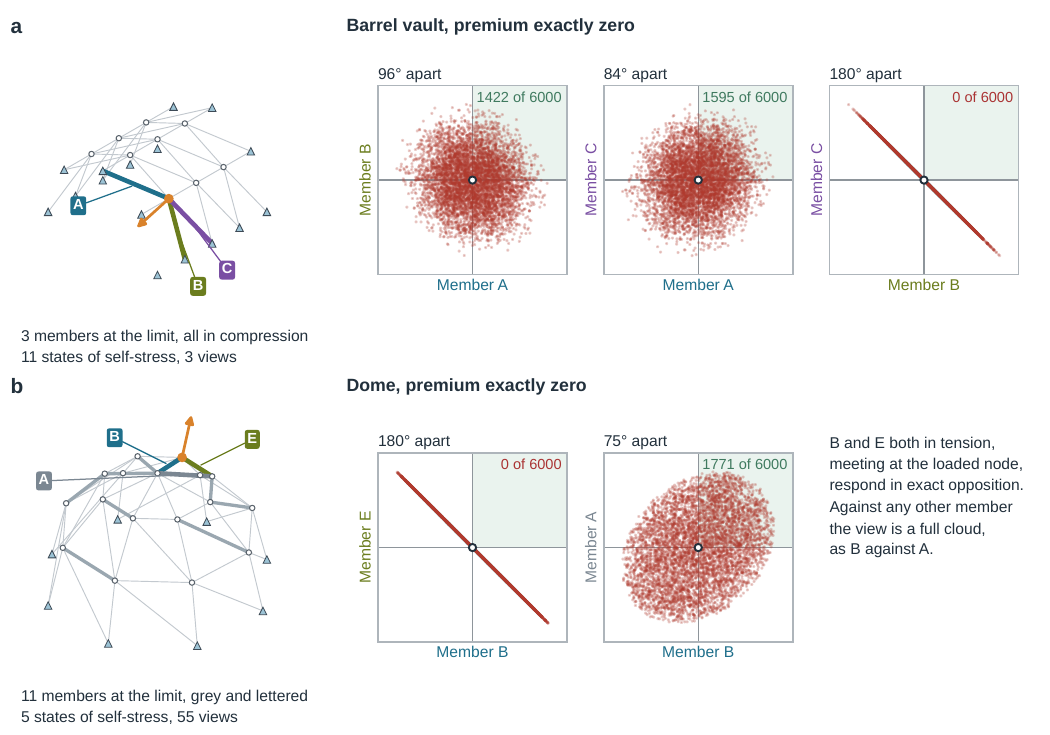}

\textbf{Supplementary Figure S6.} The block in each of the two curved lattices. Each row is
one loaded node of the sweep of Figure 7 at which the premium is exactly
zero, drawn at the load case which governs the actuated design with the
loaded node in orange and the members at the limit lettered. The views
are as in Supplementary Figure S3: each point is one of 6000 candidate
self-stresses, placed by how much it changes the two members named on
the axes, and the shaded quarter is where both fall. (a) The barrel
vault with three members at the limit, all in compression. Members B and
C respond in exact opposition to every self-stress, so the candidates
lie on one line which never enters the shaded quarter, and no candidate
lowers all three. B runs along the hoop from the loaded node and C is
the diagonal from it. (b)
The dome with eleven members at the limit and five states of self-stress.
The certificate of Eq. (7) rests on two of them alone, members B and E, which are the ties from the loaded node to the apex and to its ring neighbour, and those two respond in exact opposition. Paired with any of the other nine members, member B gives a full cloud with candidates in the shaded quarter, and the panel of B against A shows one such pair. The pair B, E is therefore the block and the other nine do not need to be examined. In both curved lattices the block is a coupled pair meeting at the
loaded node, the second block of the self-stress test. The node is sized against its own best fixed design at the matched material budget, with Euler member capacity at slenderness 91 and the member sizes frozen under actuation.

\includegraphics[width=1\textwidth,height=\textheight]{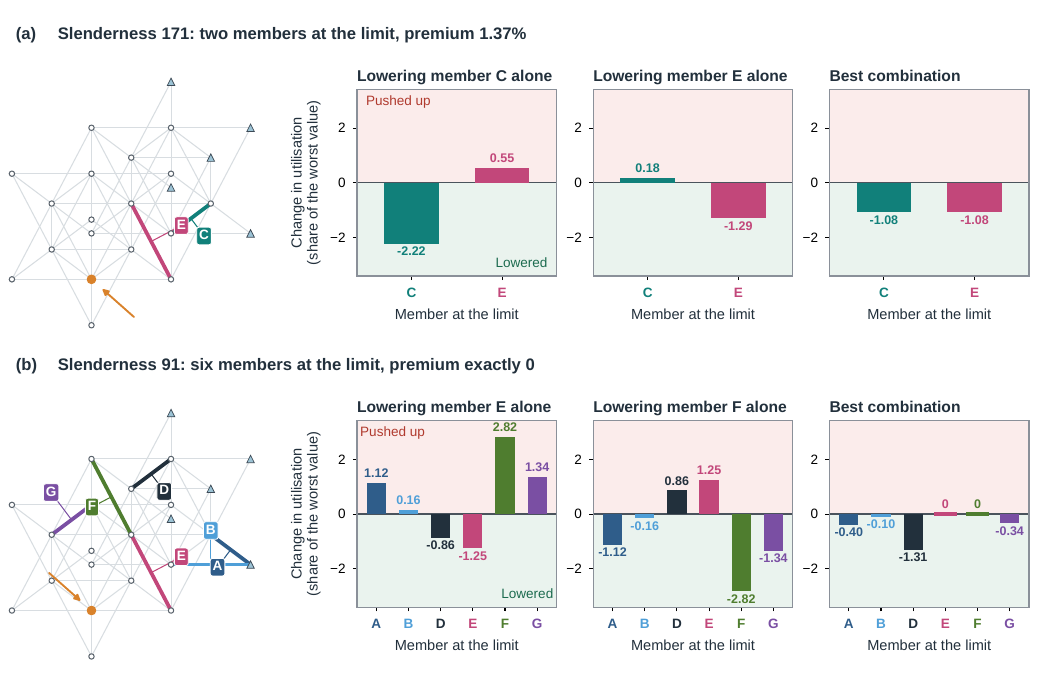}

\textbf{Supplementary Figure S7.} Slenderness can remove a zero caused
by coupled members. The same octet two-cell is loaded alone at one node
(orange) in twenty directions, and each row draws the load case which governs at that slenderness, which is a different direction in the two rows, one direction at slenderness 171 and its exact reverse at 91, with the members at the limit lettered and
coloured as on the axes; a member keeps its letter and colour in both
rows, so E is at the limit at both slendernesses. (a) At slenderness 171
two members are at the limit together and one self-stress lowers both, so
the premium is 1.37\% and the load factor rises to 1.014. (b) At
slenderness 91 six members are at the limit together, and members E and F
are coupled in exact opposition, so the self-stress which lowers one
raises the other. The best combination lowers the other four members but
leaves E and F where they are, so the load factor stays at 1.00 and the
premium is exactly zero. The self-stress space is the same in both rows
because it is set by the geometry of the structure, and what changes
with slenderness is which members are at the limit together and how their
capacities weigh them. Bars give the change in utilisation of each member at the limit, as a share of the worst value, for the self-stress which lowers the named member most (left and middle) and for the combination which lowers the worst of them most (right). Each self-stress is a mix of the ten self-stress states with every weight held between $-1$ and 1. The bound applies to the weights and not to the bars, so a bar can exceed 1 in size. Since any multiple of a self-stress is also a self-stress, only the signs and ratios of the bars matter, and the size of a bar is therefore not the premium. The premium is set by how far these members can fall before a member below the limit rises to meet them. Across all eighteen free nodes of the octet and five
slendernesses from 24 to 653, eleven nodes change between zero and
positive, and no zero caused by a member of zero reach becomes positive.
At each slenderness the node is sized against its own best fixed design at the matched material budget, with Euler member capacity and the member sizes frozen under actuation.

\includegraphics[width=1\textwidth,height=\textheight]{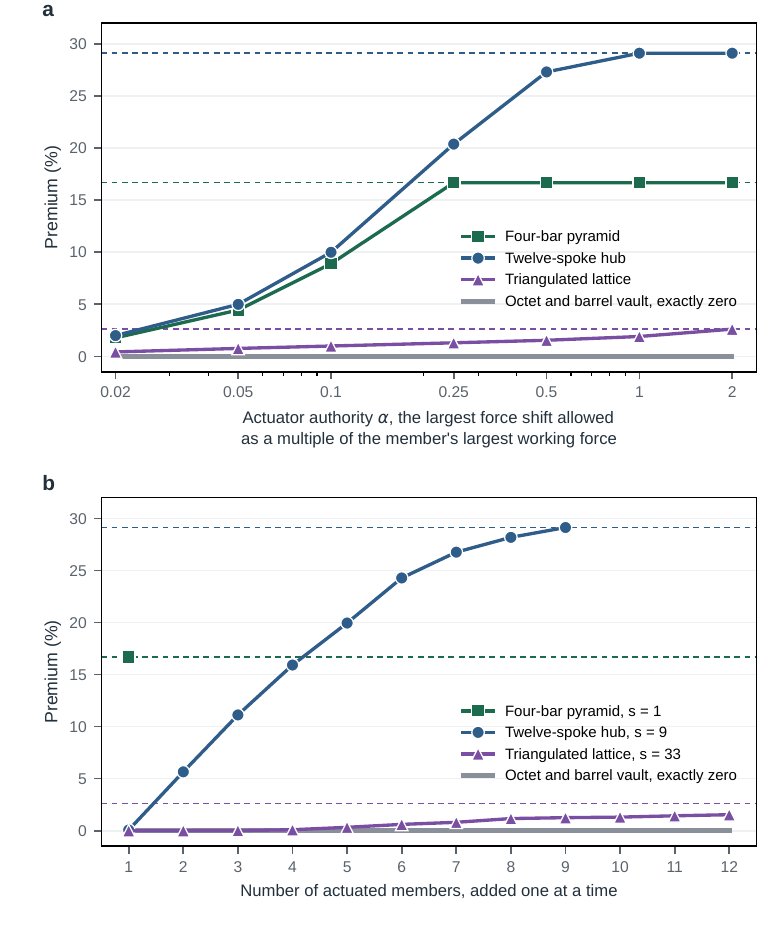}

\textbf{Supplementary Figure S8.} The zeros do not depend on the
actuator. (a) Premium against the authority of the actuators, which is
the largest force shift any actuated member may take as a multiple of
its own largest working force under the twenty load directions, on a log
scale from 0.02 to 2. (b) Premium against the number of actuated
members, with the members added one at a time in the order which raises
the premium most and \(s\) the number of self-stress states of each
structure. The octet and the barrel vault stay at exactly zero at
every bound and every count, drawn as the grey line along the axis,
while the hub, the pyramid and the triangulated lattice rise towards the value they attain with unbounded actuators on every member, drawn dashed.
The hub needs an authority near 1 and nine of its twelve members actuated before it attains its full premium, the pyramid has one state and attains its full premium with its first actuator, and the lattice has 1.53\% of its
2.62\% with twelve of its 117 members actuated. Euler member capacity,
median slenderness 91, member sizes frozen at the best
fixed design, twenty load directions.

\end{document}